\documentclass[twocolumn,showkeys,noprintnumbers,amssymb,aps,superscriptaddress,prr,longbibliography]{revtex4-2}
\usepackage{graphicx}
\usepackage{float}
\usepackage{dcolumn}
\usepackage{bm}
\usepackage{color}
\usepackage{xcolor}
\usepackage[T1]{fontenc}
\usepackage{amsmath}
\usepackage[breaklinks,colorlinks,bookmarks=false,citecolor=blue,linkcolor=red,urlcolor=blue]{hyperref}

\begin{document}

\title{Magnetization plateaus, spin-canted orders and field-induced transitions in a spin-1/2
Heisenberg antiferromagnet on a distorted diamond-decorated honeycomb lattice}
\author{Katar\'ina Kar{l}'ov\'a}
\email{katarina.karlova@upjs.sk}
\affiliation{Institute of Physics, Faculty of Science, P. J. \v{S}af\'{a}rik University, Park Angelinum 9, 04001 Ko\v{s}ice, Slovakia}
\author{Jozef Stre\v{c}ka}
\affiliation{Institute of Physics, Faculty of Science, P. J. \v{S}af\'{a}rik University, Park Angelinum 9, 04001 Ko\v{s}ice, Slovakia}

\date{\today}

\begin{abstract}
We investigate the spin-1/2 Heisenberg antiferromagnet on a distorted diamond-decorated honeycomb lattice in an external magnetic field. By combining density-matrix renormalization group, sign-problem-free quantum Monte Carlo in a mixed dimer–monomer basis, exact diagonalization, and an effective lattice-gas approach, we determine the ground-state phase diagram and analyze the finite-temperature magnetization process. The model hosts a rich variety of frustration-induced quantum phases including a quantum ferrimagnetic phase of Lieb–Mattis type, a quantum ferromagnetic phase, a spin-canted phase, a monomer–dimer phase,  a dimer–tetramer liquid, a dimer-tetramer solid, and two distinct one-dimensional-crossover phases of ferromagnetic and ferrimagnetic character. Depending on the lattice distortion, we identify robust magnetization plateaus at 0, 1/4, 1/2, and 3/4 of the saturation magnetization originating from competing local dimer and tetramer singlets. Finite-temperature QMC data reveal how thermal fluctuations progressively smear the plateau structure, while the effective lattice-gas description reliably captures the corresponding low-temperature behavior.

\end{abstract}

\maketitle

\section{Introduction}
Frustrated quantum Heisenberg antiferromagnets continue to attract considerable attention due to the wealth of unconventional quantum phases emerging from the competition between exchange interactions and quantum fluctuations \cite{diep04,lacr11}. Depending on the lattice geometry and coupling hierarchy, such systems may host quantum spin liquids, valence-bond solids, spin-canted states, ferrimagnetic phases of the Lieb–Mattis type, as well as various partially ordered or dimensional-crossover regimes \cite{bale10,stre17,stre22,karl22}. A hallmark of these phases in a magnetic field is the appearance of fractional magnetization plateaus \cite{hone04,rich04}, which encode the underlying quantum order and often reflect nontrivial commensurability conditions imposed by frustration \cite{rich02}. From a theoretical perspective, understanding such phases remains highly challenging, as frustrated two-dimensional spin systems typically lie beyond the reach of controlled analytical methods and are notoriously difficult for numerical approaches due to the sign problem or severe finite-size effects \cite{signProblem,sign2,schn18}.

Two-dimensional frustrated Heisenberg antiferromagnets in an applied magnetic field can moreover exhibit unconventional field-driven phase transitions, including transitions governed by emergent degrees of freedom absent in the microscopic Hamiltonian \cite{derz07,derz06,derz11,derz15}. A paradigmatic example is the spin-1/2 Heisenberg antiferromagnet on the kagom\'e lattice, where the lowest magnon branch becomes dispersionless and the system maps, in the high-field regime, onto a hard-hexagon gas on the triangular lattice \cite{zhit05,zhit04}. This mapping, rooted in the theory of localized magnons, predicts an order–disorder transition belonging to the universality class of the classical hard-hexagon model. Such flat-band–induced localization phenomena, originally formulated in the context of exact localized many-magnon states, also appear in other frustrated two-dimensional geometries hosting dispersionless one-magnon branches \cite{derz07,derz06,derz11,derz15}. Importantly, magnon crystals and associated fractional magnetization plateaus have been detected in real quantum magnets such as Cd-kapellasite offering an experimental realization of the spin-1/2 Heisenberg antiferromagnet on a kagom\'e lattice, thereby firmly establishing the relevance of flat-band physics as a unifying concept across frustrated quantum materials \cite{okum19}.

At the same time, significant progress has been made in developing quantum Monte Carlo (QMC) techniques capable of circumventing the sign problem in selected classes of frustrated two-dimensional Heisenberg models \cite{webe1,webe2,webe3,alet16,stap18}. By performing simulations in an appropriate dimer or trimer basis—rather than in the conventional local spin-$S^{z}$ basis—the sign problem can be eliminated even in the presence of geometric frustration. This strategy has enabled large-scale, sign-free QMC studies of a variety of frustrated two-dimensional systems including the Shastry–Sutherland lattice \cite{webe1}, fully frustrated bilayer \cite{alet16,stap18} and trilayer \cite{webe2,webe3} antiferromagnets, and diamond-decorated square lattices \cite{caci23,karl24}, where the Hamiltonian admits a natural decomposition into few-site clusters. These methodological developments provide a powerful numerical framework for addressing thermodynamics and field-induced phenomena in systems that would otherwise be inaccessible \cite{webe1,webe2,webe3,alet16,stap18}.

In this work we focus on a frustrated two-dimensional quantum Heisenberg antiferromagnet defined on the diamond-decorated honeycomb lattice, which represents a natural extension of the diamond-decorated square lattice studied previously \cite{mori16,hiro16,hiro17,hiro18,hiro20,caci23,karl24,dmit1}. The magnetic structure of the diamond-decorated honeycomb lattice is inspired by a metallic framework of bimetallic two-dimensional coordination polymers [\{Cu(bipn)\}$_3$Fe(CN)$_6$](ClO$_4$)$_{2}\cdot4$H$_2$O [bipn = bis(3-amino\-propyl)-amine] \cite{zhang2000} and [\{Cu(ept)\}$_3$Fe(CN)$_6$](ClO$_4$)$_2\cdot5$H$_2$O [ept = $N$-(2-amino\-ethyl)-1,3-di\-amino\-pro\-pane] \cite{travnicek2001}. We investigate the low-temperature magnetization process of this model in the presence of lattice distortion that modifies the hierarchy of the exchange couplings on the vertical and zigzag diamond units. By combining large-scale density-matrix renormalization group (DMRG) calculations, sign-problem–free QMC simulations, exact diagonalization (ED), and an analytical effective lattice-gas description derived from the framework of localized magnons, we construct the ground-state phase diagram and analyze the finite-temperature evolution of the magnetization plateaus. This multi-method approach enables us to unravel a series of unconventional quantum phases and field-induced transitions stabilized through frustration, flat-band physics, and the competing local structures of the diamond units.

\section{Heisenberg model on the diamond-decorated honeycomb lattice}
\label{model}
\begin{figure}[t!]
	\centering
	\includegraphics[width=0.9\columnwidth]{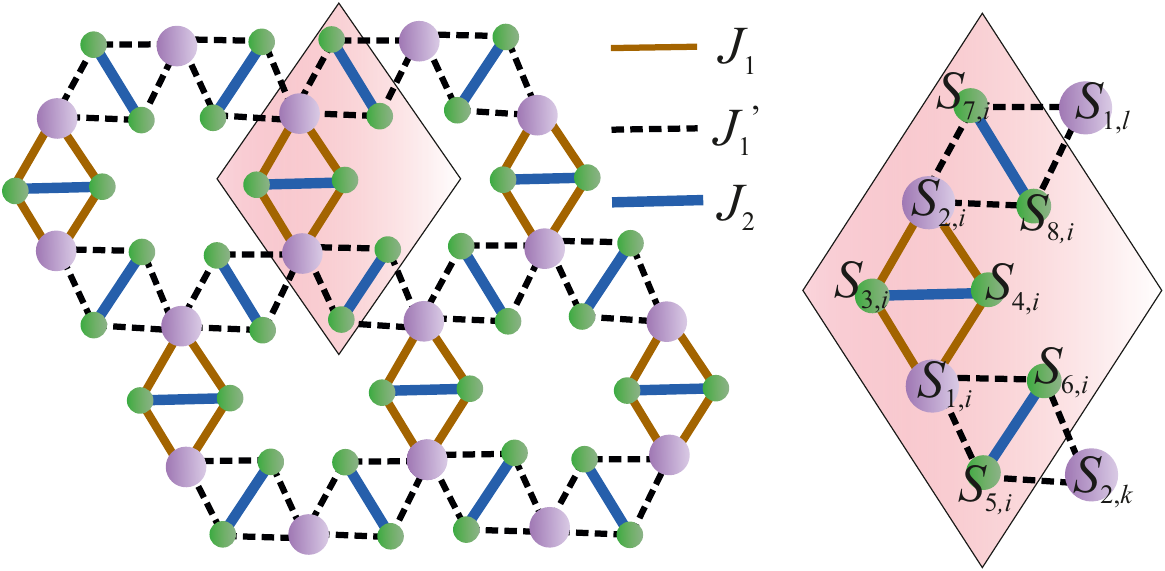}
	\caption{A schematic illustration of the spin-1/2 Heisenberg model on a distorted diamond-decorated honeycomb lattice. The coupling constants $J_1$, $J_1'$ and $J_2$ are indicated by lines of distinct colors. styles, and thicknesses. A single unit cell is highlighted by the pink rhombus. The right panel shows an enlarged view of the eight-spin cluster forming the unit cell including the site-numbering convention.}
	\label{fig:model}
\end{figure}

We consider the spin-$1/2$ Heisenberg model on a distorted diamond-decorated honeycomb lattice schematically depicted in Fig.~\ref{fig:model} and defined by the Hamiltonian
\begin{align}
\hat{\mathcal H} \! &= \! \sum_i \Big\{ J_1 (\hat{\mathbf S}_{1,i} \!+\! \hat{\mathbf S}_{2,i}) \!\cdot\! (\hat{\mathbf S}_{3,i} \!+\! \hat{\mathbf S}_{4,i}) \nonumber \\
\! &+ \! J_1' \big[(\hat{\mathbf S}_{5,i} \!+\! \hat{\mathbf S}_{6,i}) \!\cdot\! (\hat{\mathbf S}_{1,i} \!+\! \hat{\mathbf S}_{2,k})
           \!+\! (\hat{\mathbf S}_{7,i} \!+\! \hat{\mathbf S}_{8,i}) \!\cdot\! (\hat{\mathbf S}_{2,i} \!+\! \hat{\mathbf S}_{1,l})\big] \nonumber \\
\! &+ \! J_2 \big(\hat{\mathbf S}_{3,i}\!\cdot\!\hat{\mathbf S}_{4,i} \!+\! \hat{\mathbf S}_{5,i}\!\cdot\!\hat{\mathbf S}_{6,i} \!+\! \hat{\mathbf S}_{7,i}\!\cdot\!\hat{\mathbf S}_{8,i} \big) 
 \!-\! h\sum_{j=1}^8 \hat{S}_{j,i}^z \Big\}. 
\label{hamos}
\end{align}
Here, $\hat{\mathbf S}_{j,i}$ ($j=1,\dots,8$) represents a spin-$1/2$ operator assigned to the $j$-th site of the $i$-th unit cell following the site numbering convention illustrated in Fig.~\ref{fig:model}. Each unit cell consists of two monomer sites labeled with site indices $j=1,2$ and three pairs of dimer sites arranged into one vertical diamond unit ($j=3,4$) and two zigzag diamond units ($j=5-8$). Three distinct coupling constants $J_1$, $J_1'$, and $J_2$ are assumed within this frustrated spin system. The coupling constant $J_{1}$ couples monomer spins to the dimer spins on the vertical diamonds, whereas the coupling constant $J_{1}'$ couples monomer spins to the dimer spins on the zigzag diamonds. The lattice distortion is introduced through the parametrization $J_{1}' = J_{1}(1+\delta_{1})$, where $\delta_{1}$ quantifies the relative change of the coupling constant between the monomer and dimer spins on the zigzag bonds with respect to the one on the vertical bonds. A negative distortion $\delta_{1}<0$ thus weakens the coupling constant within zigzag diamond units with respect to the vertical ones $J_{1}'<J_{1}$, while a positive distortion $\delta_{1}>0$ contrarily enhances it $J_{1}'>J_{1}$. By contrast, the coupling constant $J_{2}$ represents the intradimer exchange interaction, which is assumed to be uniform for all diamond units.  Finally, the last term in Eq.~(\ref{hamos}) accounts for the Zeeman energy in an external magnetic field with the field strength given by $h=g\mu_{\rm B}B$, where $g$ is the Land\'e $g$-factor, $\mu_{\rm B}$ is the Bohr magneton, and $B$ is the applied magnetic field. Throughout this work, we set $g\mu_{\rm B}=1$ for convenience, so that $h$ directly represents the magnetic-field strength in energy units.

The spin-1/2 Heisenberg model on a distorted diamond-decorated honeycomb lattice defined by the Hamiltonian (\ref{hamos}) belongs to a special class of frustrated quantum spin systems with locally conserved quantities. In particular, the total spin of the decorating dimers defined through the composite spin operators $\hat{\mathbf S}_{34,i} = \hat{\mathbf S}_{3,i} + \hat{\mathbf S}_{4,i}$, $\hat{\mathbf S}_{56,i} = \hat{\mathbf S}_{5,i} + \hat{\mathbf S}_{6,i}$, and $\hat{\mathbf S}_{78,i} = \hat{\mathbf S}_{7,i} + \hat{\mathbf S}_{8,i}$ represent a set of conserved quantities with well-defined quantum spin numbers as their squares commute with the Hamiltonian $[\hat{\mathcal H},\hat{\mathbf S}_{34,i}^2] = [\hat{\mathcal H},\hat{\mathbf S}_{56,i}^2] = [\hat{\mathcal H},\hat{\mathbf S}_{78,i}^2] = 0$. Exploiting this property, the Hamiltonian (\ref{hamos}) of the spin-1/2 Heisenberg model on a distorted diamond-decorated honeycomb lattice can be equivalently rewritten in terms of the composite spin operators as
\begin{align}
\hat{\mathcal H} \!=\! \sum_i \!\! \Big\{ \!
& J_1 \! (\hat{\mathbf S}_{1,i} \!+\! \hat{\mathbf S}_{2,i}) \!\cdot\! \hat{\mathbf S}_{34,i} + 
\frac{J_2}{2} \! \big(\hat{\mathbf S}_{34,i}^2 \!+\! \hat{\mathbf S}_{56,i}^2 \!+\! \hat{\mathbf S}_{78,i}^2 \!-\! \frac{9}{2} \big) 
\nonumber \\
& \!+\! J_1' [(\hat{\mathbf S}_{1,i} \!+\! \hat{\mathbf S}_{2,k}) \!\cdot\! \hat{\mathbf S}_{56,i} 
\!+\! J_1' (\hat{\mathbf S}_{2,i} \!+\! \hat{\mathbf S}_{1,l}) \!\cdot\! \hat{\mathbf S}_{78,i}] \nonumber \\
& \!-\! h  (\hat{S}_{1,i}^z + \hat{S}_{2,i}^z + \hat{S}_{34,i}^z + \hat{S}_{56,i}^z + \hat{S}_{78,i}^z) \Big\}.
\label{dmrg}
\end{align}
This mapping yields an equivalent representation of the original frustrated quantum spin system, which is however free of spin frustration as the resulting Hamiltonian (\ref{dmrg}) corresponds to an unfrustrated mixed-spin Heisenberg model on a Lieb-type honeycomb lattice. Within this alternative formulation, each composite spin takes one of two possible values 0 or 1 depending on whether the corresponding decorating dimer is in a singlet or triplet state. The alternative representation of the Hamiltonian (\ref{dmrg}) not only provides a route to sign-problem-free QMC simulations in a basis spanned over all eigenstates of the decorating dimers, but it also offers direct access to several fragmented exact ground states and enables more efficient DMRG simulations. 

\section{Numerical methods}
\label{method}

In this section, we provide specific details of three numerical methods DMRG, ED, and QMC used for investigating the spin-1/2 Heisenberg model on a distorted diamond-decorated honeycomb lattice.

\begin{figure}[t!]
	\centering
	\includegraphics[width=\columnwidth]{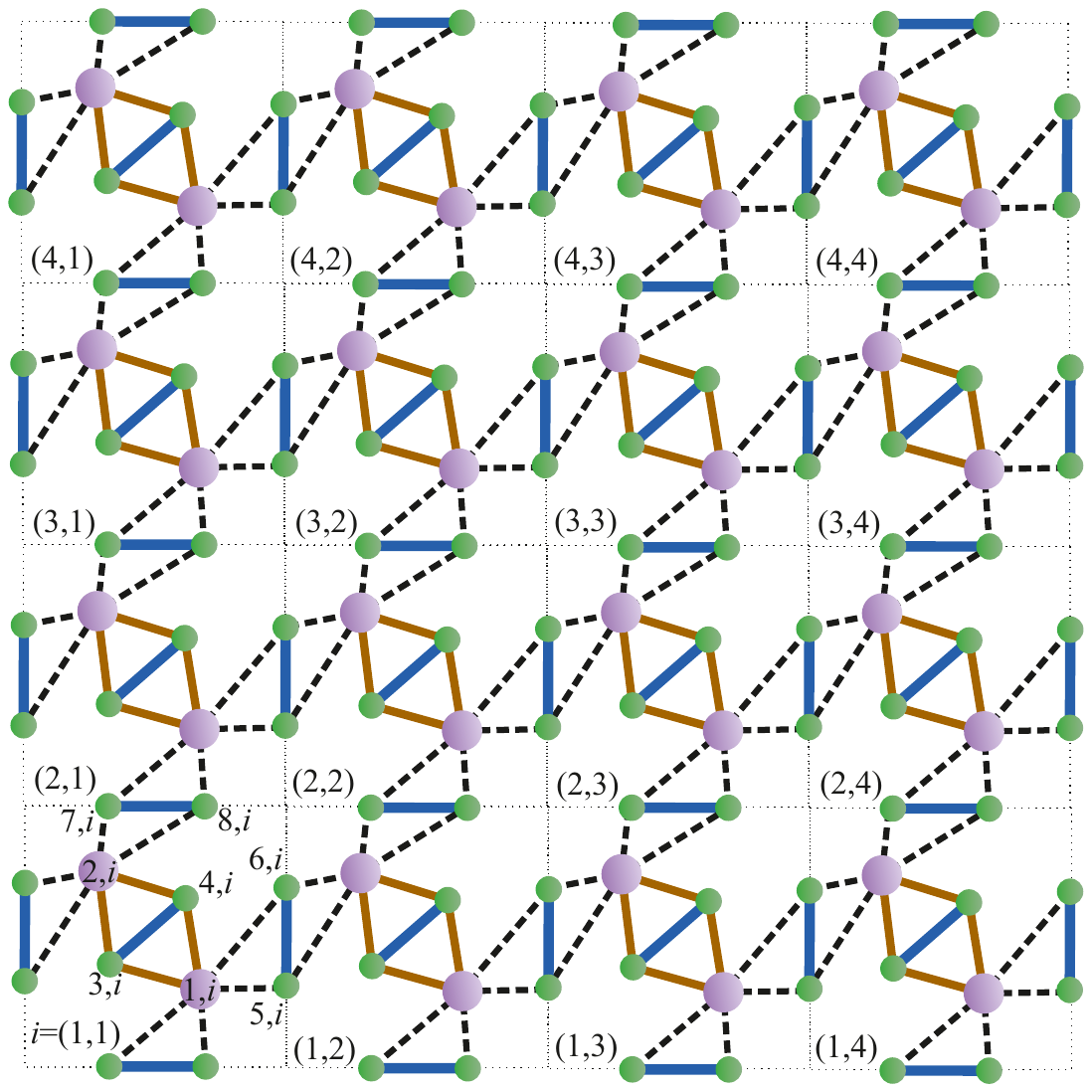}
	\vspace{-3.5cm} 
	\caption{Finite-size cluster of a distorted diamond-decorated honeycomb lattice with linear system size $L=4$ (16 unit cells) used for DMRG and QMC simulations. The numbers in parentheses denote unit-cell coordinates (row, column), whereas toroidal boundary conditions were applied.}
	\label{fig:4x4}
\end{figure}

\subsection{Density-matrix renormalization group}

Exploiting the frustration-free representation of the Hamiltonian (\ref{dmrg}), we performed DMRG simulations of the mixed-spin Heisenberg model on a Lieb-type honeycomb lattice using an adapted version of the \texttt{dmrg} routine from the Algorithms and Libraries for Physics Simulations (ALPS) project \cite{baue11}. Within the DMRG calculations corresponding to a finite lattice with linear size $L=4$ (16 unit cells) shown in Fig. \ref{fig:4x4}, we obtained lowest-energy eigenstate in each sector with a fixed $z$-component of the total spin $S_T^z$ by systematically considering all admissible combinations of composite quantum spin numbers. To guarantee reliable convergence, we retained up to 1200 states in the DMRG procedure and carried out 10 sweeps under toroidal boundary conditions, which proved sufficient to keep truncation errors at an acceptably low level throughout the entire parameter range investigated.

\subsection{Exact diagonalization}
To validate the zero-temperature magnetization curves and the ground-state phase diagram obtained from the DMRG results, we performed ED calculations for the spin-1/2 Heisenberg model on a finite distorted diamond-decorated honeycomb lattice composed of $2\times2$ unit cells corresponding to the total number of $N=32$ spins. For this purpose, we employed the \texttt{sparsediag} routine from the \textsc{ALPS} project \cite{baue11}, which provides an efficient implementation of iterative diagonalization of sparse matrices based on the Lanczos algorithm by constructing a Krylov subspace generated through repeated applications of the Hamiltonian on a trial vector and iteratively converging to the lowest-energy eigenstates. It is noteworthy that these ED calculations were carried out for the original formulation of the model given by the Hamiltonian (\ref{hamos}) rather than its frustration-free representation (\ref{dmrg}).

\subsection{Quantum Monte Carlo}
The finite-temperature properties of the spin-1/2 Heisenberg model on a distorted diamond-decorated honeycomb lattice were investigated primarily by means of QMC simulations, which enable access to sufficiently large system sizes over a broad range of temperatures such as $4\times 4$ unit cells comprising a total number of $N=128$ spins as schematically illustrated in Fig. \ref{fig:4x4}. For this purpose, we employed the stochastic series expansion (SSE) formulation with directed-loop updates, which provides unbiased and numerically efficient sampling for quantum spin models. A direct application of SSE in the conventional local $S^{z}$ basis would, however, suffer from the usual sign problem once geometric frustration is present. In our model, the frustration originates from the antiferromagnetic intradimer interaction $J_{2}$ and a reformulation of the local basis is therefore required.

To circumvent this difficulty, we follow a strategy previously developed for related frustrated quantum spin systems and naturally connected to the frustration-free (Lieb-type) representation of the Hamiltonian (\ref{dmrg}): spins connected by the frustrating $J_{2}$-bonds are treated in the eigenbasis of the dimer Hamiltonian, whereas all monomeric spins remain expressed in the standard $S^{z}$ basis. This mixed representation leads to a five-site local basis for each unit cell consisting of two monomer spins and three composite dimer spins, in which all matrix elements of the SSE operator string are strictly non-negative. Consequently, the Monte Carlo sampling is completely free of the sign problem despite the geometric frustration. Technical details of this computational procedure can be found in Refs.~\cite{caci23,karl24}.

\section{Ground-state properties}
\label{gspd}
\begin{figure*}[t!]
	\centering
	\includegraphics[width=1\columnwidth]{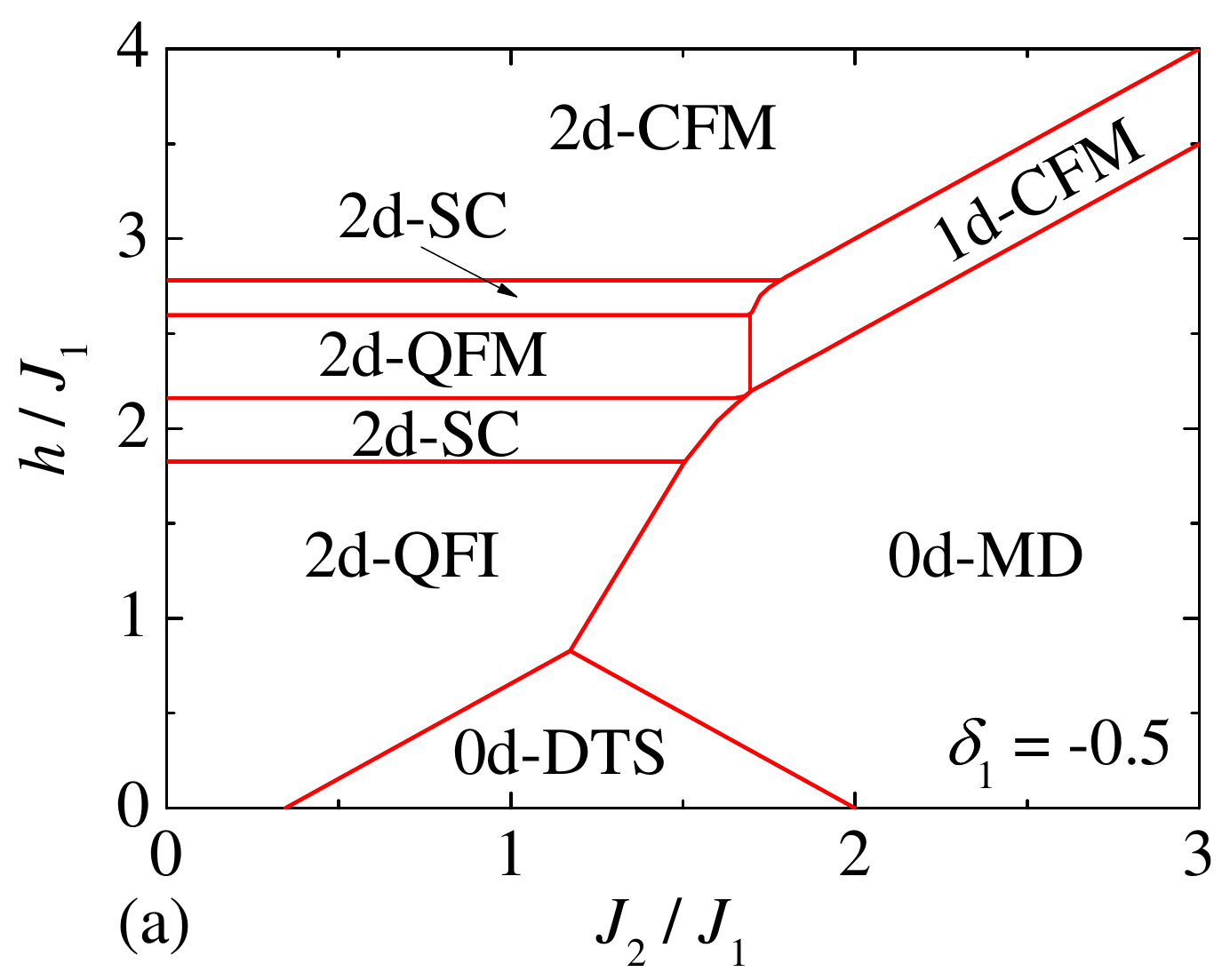}\hspace{-0.2cm}
	\includegraphics[width=1\columnwidth]{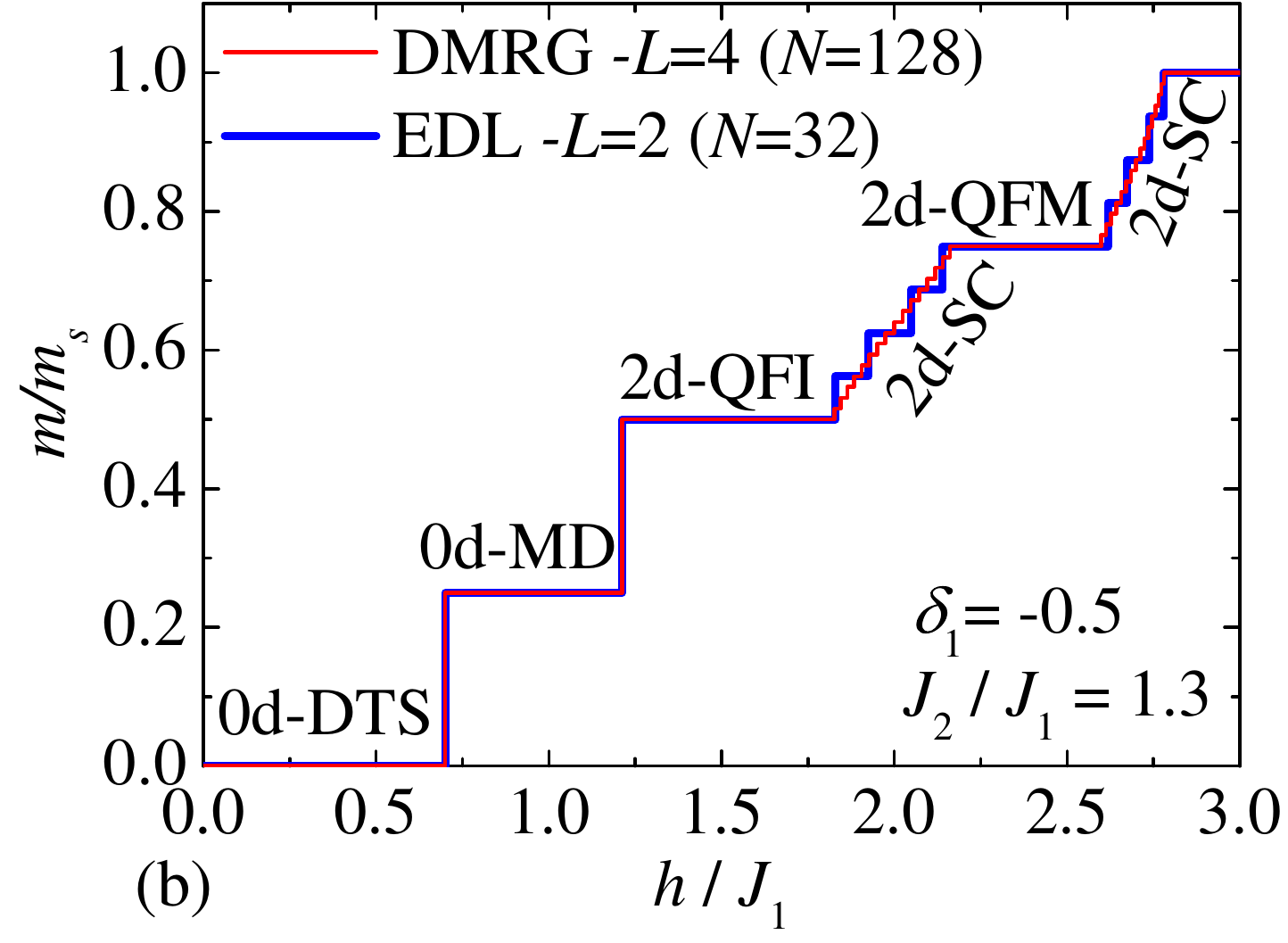}
		\includegraphics[width=1\columnwidth]{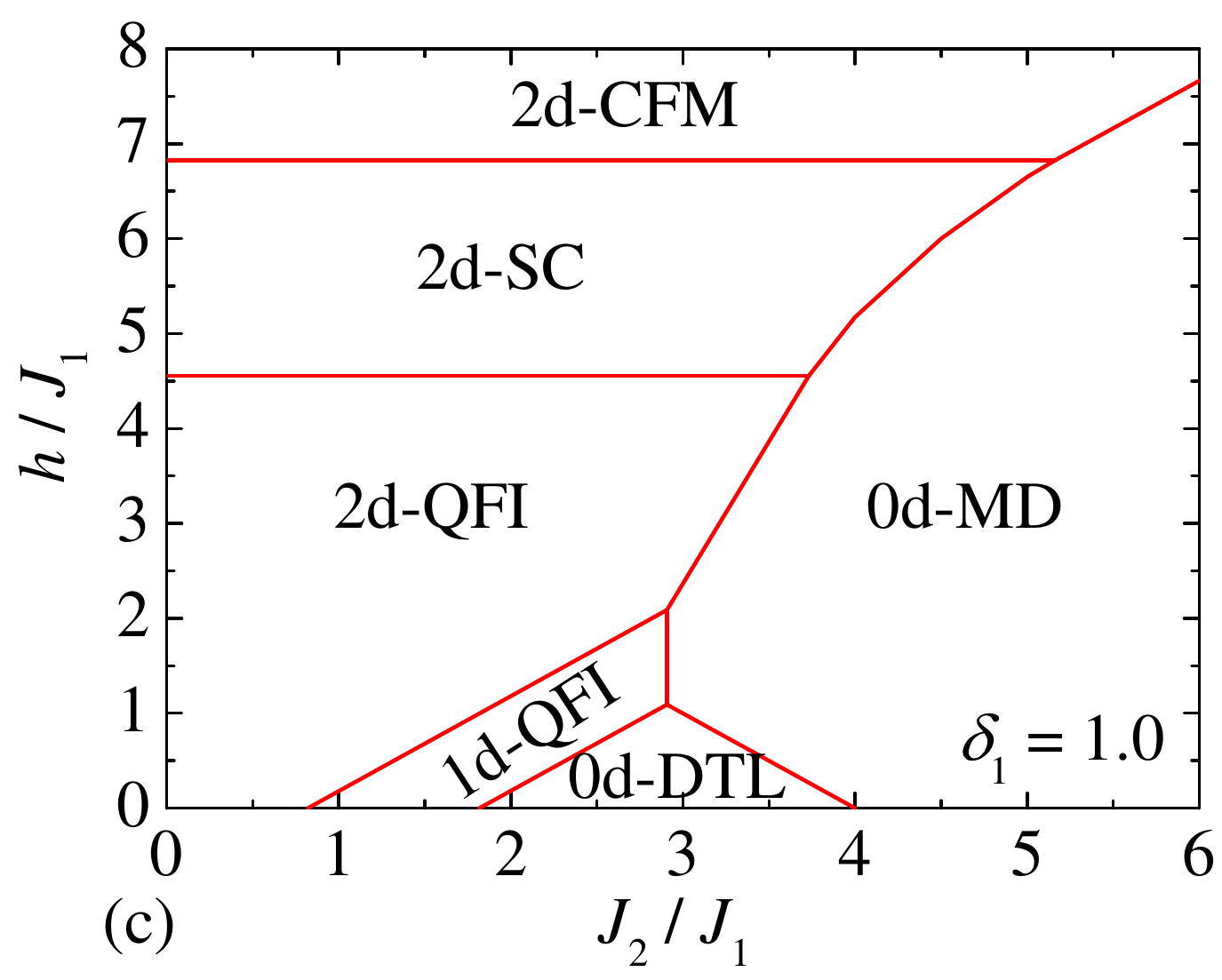}\hspace{-0.2cm}
	\includegraphics[width=1\columnwidth]{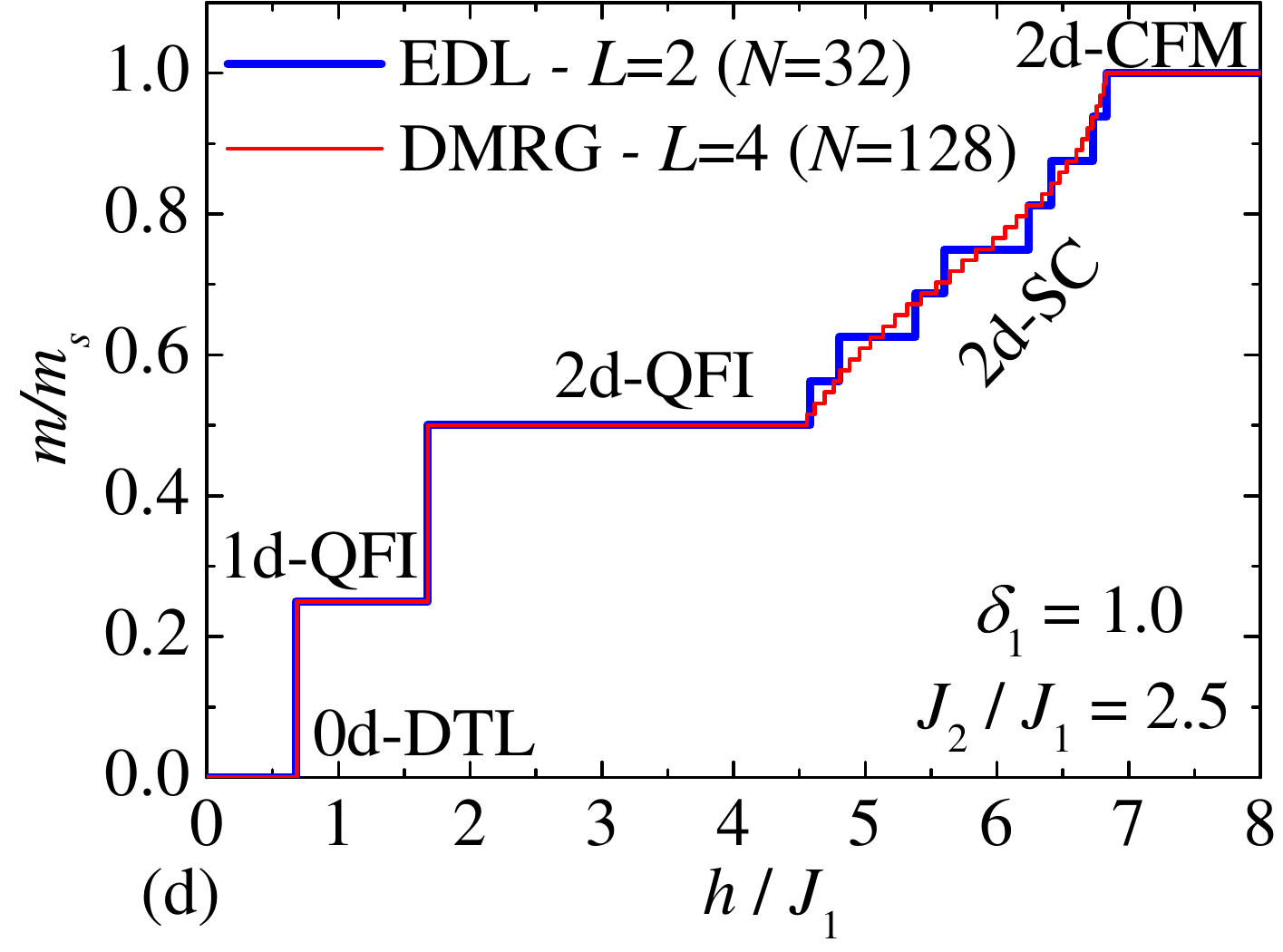}
	\caption{Ground-state phase diagrams of the spin-1/2 Heisenberg model on the diamond-decorated honeycomb lattice in the $J_{2}/J_{1}$–$h/J_{1}$ plane for two representative values of the distortion parameter $\delta_{1}=-0.5$ (a) and $1.0$ (c) as obtained from DMRG simulations for a linear system size $L=4$ (128 spins). Panels (b) and (d) illustrate zero-temperature magnetization curves for two selected values of the coupling ratio $J_2/J_1$ as obtained from DMRG simulations for $L=4$ and cross-validated by ED using the Lanczos algorithm for $L=2$.}
	\label{fig:GSPD}
\end{figure*}

\begin{figure*}[t!]
\centering
\includegraphics[width=1\textwidth]{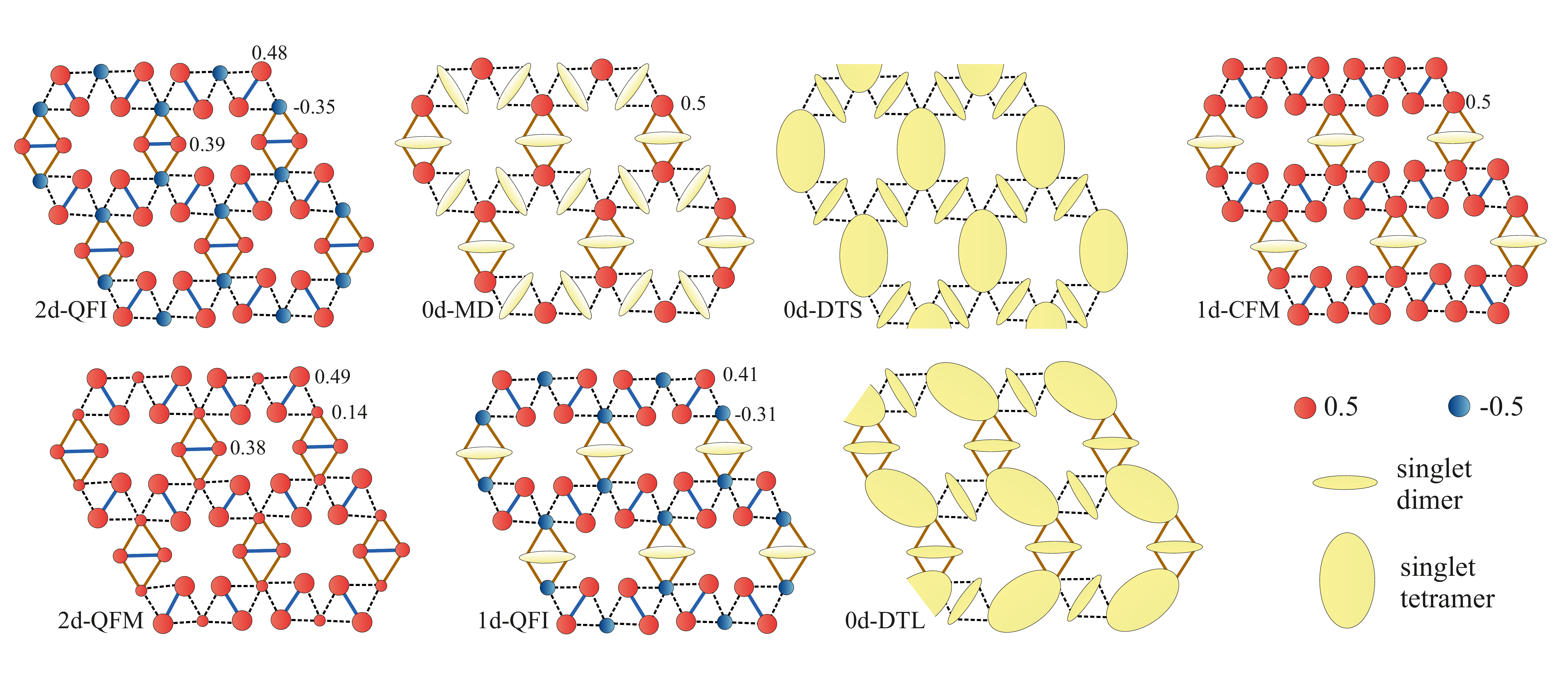}\hspace{-0.2cm}
\vspace{-0.75cm}
\caption{Ground states of the spin-1/2 Heisenberg model on the diamond-decorated honeycomb lattice. Singlet dimers and singlet tetramers are shown by small and large yellow ovals, while red (blue) circles denote sites with positive (negative) local magnetization. The radius of each circle is proportional to the magnitude of the local magnetization with $m = \pm 0.5$ representing the extremal on-site values. Numbers shown in the panels indicate specific numerical values of the local magnetizations rounded to two decimal places as obtained from DMRG simulations for a finite system of linear size $L=4$ ($N=128$ spins). The notation of individual ground states: 2d quantum ferrimagnetic (2d-QFI) phase, 2d quantum ferromagnetic (2d-QFM) phase, 1d classical ferromagnetic (1d-CFM) phase, 1d quantum ferrimagnetic (1d-QFI) phase, 0d monomer-dimer (0d-MD) phase, 0d dimer-tetramer solid (0d-DTS) phase, and 0d dimer-tetramer liquid (0d-DTL) phase.}
\label{fig:fazy}
\end{figure*}

We have determined the ground-state phase diagram of the spin-1/2 Heisenberg model on the diamond-decorated honeycomb lattice in the $J_{2}/J_{1}$–$h/J_{1}$ plane, which is depicted in Fig.~\ref{fig:GSPD} for two representative values of the distortion parameter $\delta_{1}=-0.5$ and $1.0$. The phase boundaries were obtained from DMRG calculations for a system of linear size $L=4$ corresponding to 128 spins. For the negative distortion parameter $\delta_{1}=-0.5$ shown in Fig.~\ref{fig:GSPD}(a), the system exhibits a rich variety of quantum phases including the 2d quantum ferrimagnetic (2d-QFI) phase, the 2d spin-canted (2d-SC) phase, the 0d monomer–dimer (0d-MD) phase, the 0d dimer–tetramer solid (0d-DTS) phase, the 2d quantum ferromagnetic (2d-QFM) phase, and the 1d classical ferromagnetic (1d-CFM) phase. The qualitative nature of these phases is illustrated in Fig.~\ref{fig:fazy}.

It follows from Fig.~\ref{fig:fazy} that the 2d-QFI phase exhibits a ferrimagnetic character: the dimer triplets tend to align parallel to the external magnetic field, whereas the monomer spins align antiparallel with respect to it. The local magnetizations reveal a substantial quantum reduction of the magnetic moments, which varies significantly among the inequivalent lattice sites and strongly depends on both the type and magnitude of the distortion. The numerical values of the local magnetizations quoted in Fig.~\ref{fig:fazy} for the 2d-QFI phase correspond to the case of negative distortion  $\delta_{1}=-0.5$ where the dimer triplets on the zigzag diamonds undergo a smaller quantum reduction than those on the vertical diamonds, while the opposite trend is observed for the positive distortion $\delta_{1}>0$. 

By contrast, the 0d-MD phase emerging in the highly frustrated regime consists of isolated dimer singlets, which render the monomer spins effectively free (paramagnetic). Consequently, the monomer spins polarize along the field direction for any nonzero magnetic field. The designation 0d-MD reflects the zero-dimensional nature of this phase in the sense of the absence of quantum correlations other than those emergent within dimer singlets. Because all diamond units host the dimer singlets, the system undergoes a complete fragmentation and the 0d-MD ground state factorizes into a direct product of independent dimer singlets and monomer spins fully polarized along the field direction 
\begin{eqnarray}
\!\!\!&&\!\!\! |{\rm 0d\!-\!MD}\rangle = \prod_i \Big[|\!\uparrow_{1,i}\rangle |\!\uparrow_{2,i}\rangle \otimes \frac{1}{\sqrt{2}}
          \big(|\!\uparrow_{3,i}\downarrow_{4,i}\rangle \!-\! |\!\downarrow_{3,i}\uparrow_{4,i}\rangle\big) \nonumber \\
&\otimes& \frac{1}{\sqrt{2}}\big(|\!\uparrow_{5,i}\downarrow_{6,i}\rangle \!-\! |\!\downarrow_{5,i}\uparrow_{6,i}\rangle\big) 
\otimes \frac{1}{\sqrt{2}}\big(|\!\uparrow_{7,i}\downarrow_{8,i}\rangle \!-\!|\!\downarrow_{7,i}\uparrow_{8,i}\rangle\big)\Big], \nonumber \\
\end{eqnarray}
which explicitly illustrates the factorized zero-dimensional nature of the phase. 

Between the 2d-QFI and 0d-MD  phases, one encounters the 0d-DTS characterized by singlet tetramers formed uniquely on all vertical diamond units together with dimer singlets residing on all zigzag diamond units as illustrated in Fig.~\ref{fig:fazy}. Similar to the 0d-MD phase, the presence of local dimer and tetramer singlets thus leads to another completely fragmented ground state. This justifies its classification as a zero-dimensional phase, since the respective eigenvector factorizes into a direct product of independent singlet-tetramer and singlet-dimer states 
\begin{eqnarray}
|\!\!\!&&\!\!\! \mathrm{0d\!-\!DTS}\rangle \!=\! \prod_i \Big[ \frac{1}{\sqrt{3}} \Big(|\!\uparrow_{1,i}\downarrow_{2,i}\uparrow_{3,i}\downarrow_{4,i}\rangle 
\!+\! |\!\downarrow_{1,i}\uparrow_{2,i}\downarrow_{3,i}\uparrow_{4,i}\rangle \Big) \nonumber \\
&& \qquad \qquad - \frac{1}{\sqrt{12}} \Big(|\!\uparrow_{1,i}\uparrow_{2,i}\downarrow_{3,i}\downarrow_{4,i}\rangle \!+\! |\!\uparrow_{1,i}\downarrow_{2,i}\downarrow_{3,i}\uparrow_{4,i}\rangle \nonumber \\
&& \qquad \qquad \qquad \qquad + |\!\downarrow_{1,i}\uparrow_{2,i}\uparrow_{3,i}\downarrow_{4,i}\rangle \!+\! |\!\downarrow_{1,i}\downarrow_{2,i}\uparrow_{3,i}\uparrow_{4,i}\rangle \Big) \nonumber \\
&\otimes& \frac{1}{\sqrt{2}}\big(|\!\uparrow_{5,i}\downarrow_{6,i}\rangle \!-\!|\!\downarrow_{5,i}\uparrow_{6,i}\rangle\big) 
\otimes \frac{1}{\sqrt{2}}\big(|\!\uparrow_{7,i}\downarrow_{8,i}\rangle \!-\! |\!\downarrow_{7,i}\uparrow_{8,i}\rangle\big) \Big].
\nonumber \\
\end{eqnarray}

At higher magnetic fields, the competition between the magnetic spin frustration and the tendency to align towards the external magnetic field gives rise to three additional exotic ground states. In the highly frustrated regime and sufficiently strong magnetic fields, one detects the 1d-CFM phase characterized by a peculiar coexistence of dimer singlets on all vertical diamond units and fully polarized spin-1/2 chains running along the zigzag diamond units 
\begin{eqnarray}
|\mathrm{1d\!-\!CFM}\rangle \!=\! \prod_i \Big[ |\!\uparrow_{1,i}\uparrow_{2,i}\rangle 
&\otimes& \frac{1}{\sqrt{2}}\big(|\!\uparrow_{3,i}\downarrow_{4,i}\rangle \!-\! |\!\downarrow_{3,i}\uparrow_{4,i}\rangle\big) \nonumber \\
&\otimes& |\!\uparrow_{5,i}\uparrow_{6,i}\rangle \otimes |\!\uparrow_{7,i}\uparrow_{8,i}\rangle \Big].
\end{eqnarray}
The dimer singlets on the vertical diamond units repeatedly indicate a dimensional reduction due to a fragmented structure of the 1d-CFM ground state, which accordingly retains a one-dimensional classical ferromagnetic spin alignment along the zigzag diamond chains. 

In a less frustrated parameter region and under sufficiently high magnetic fields, the system undergoes a field-driven quantum phase transition from the 2d-QFI phase to the 2d-SC phase. The 2d-SC phase is a gapless critical state with a continuous excitation spectrum, which produces a smooth continuous rise of the magnetization due to a gradual spin canting towards the field direction. This behavior is clearly manifested in the zero-temperature magnetization curve displayed in Fig.~\ref{fig:GSPD}(b) for one representative value of the coupling ratio $J_2/J_1=1.3$. Remarkably, the additional gapped 2d-QFM phase is embedded into the critical region of the 2d-SC phase, which gives rise to a pronounced intermediate 3/4-plateau in a zero-temperature magnetization curve. Although the 2d-QFM phase has no classical analogue, its microscopic structure shown in Fig.~\ref{fig:fazy} reveals a predominantly ferromagnetic order with a substantial quantum reduction of the local magnetization that is most pronounced on the monomer spins acquiring the smallest value $m_{1-2} \approx 0.14$. Despite the triplet character of all dimers, they also undergo a sizable quantum reduction of local magnetic moments $m_{3-4} \approx 0.38$ on the vertical diamond units unlike the dimer spins on the zigzag diamond units approaching near-saturated value $m_{5-8} \approx 0.49$. 

A great diversity of quantum ground states described above is directly reflected in a remarkable zero-temperature magnetization curve, which involves a sequence of magnetization plateaus, jumps, and field-driven quantum phase transitions associated with the underlying quantum phases. The zero-magnetization plateau in Fig.~\ref{fig:GSPD}(b) corresponds to the gapped 0d-DTS phase, which terminates at a discontinuous field-driven transition into the 0d-MD phase manifested as the intermediate 1/4-plateau. A subsequent magnetization jump renders a second discontinuous field-induced transition from the 0d-MD phase to the 2d-QFI phase represented by the intermediate 1/2-plateau. Four additional continuous field-driven quantum phase transitions relate to the emergence or disappearance of the 2d-SC and 2d-QFM phases. A comparison of ED results for smaller spin cluster with DMRG data for larger systems reveals that the step-like character of the magnetization curve within the field region corresponding to the 2d-SC phase gradually evolves into a smooth continuous magnetization curve in the thermodynamic limit. Although the field range associated with the 2d-QFM phase becomes slightly narrower with increasing system size, the respective intermediate 3/4-plateau remains preserved even in the thermodynamic limit as evidenced by a finite-size analysis. 
 
\begin{figure}[t!]
	\centering
	\includegraphics[width=1\columnwidth]{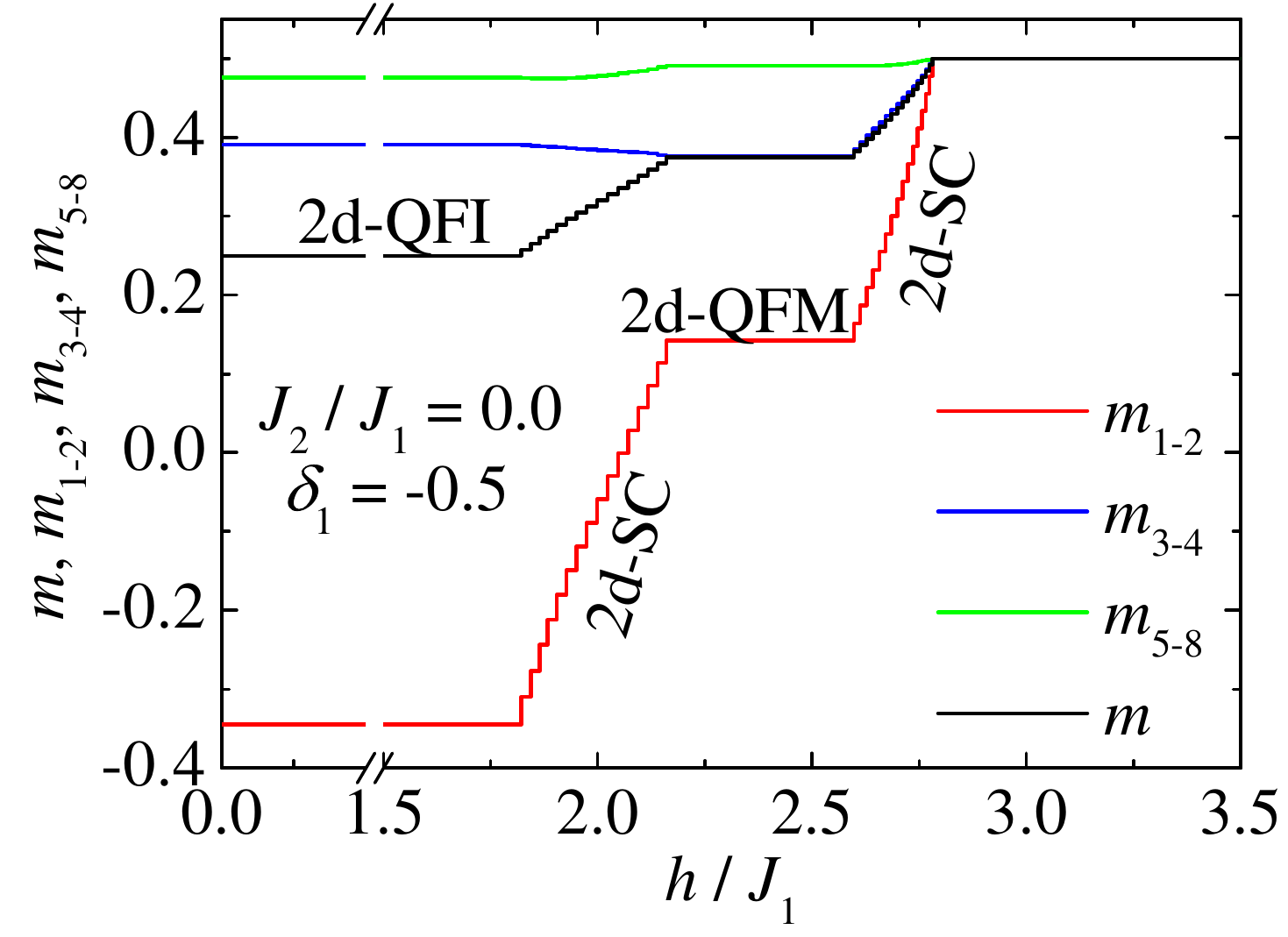}\hspace{-0.2cm}
	\vspace{-0.45cm}
	\caption{Field dependence of the total magnetization $m$ is plotted together with the local magnetizations of the monomer spins ($m_{1\text{--}2}$), 
	the dimer spins on the vertical diamonds ($m_{3\text{--}4}$), and the dimer spins on the zigzag diamonds ($m_{5\text{--}8}$) for $J_{2}/J_{1}=0$ and $\delta_{1}=-0.5$. 
  For better clarity, there is a break on the horizontal field axis within the low-field region, where all depicted magnetizations remain constant.}
	\label{fig:Lokalne}
\end{figure}

To gain deeper insight into the field-induced evolution of the magnetization, the total magnetization is plotted in Fig.~\ref{fig:Lokalne} together with the local magnetizations of the monomer spins, the dimer spins on the vertical diamond units, and the dimer spins on the zigzag diamond units for $J_2/J_1=0.0$ and $\delta_1=-0.5$. In the low-field regime, the local magnetizations confirm the ferrimagnetic character of the 2d-QFI phase, where the monomer spins exhibit a small negative polarization in contrast to the dimer spins predominantly oriented into the magnetic field. The most intriguing evolution of the magnetization occurs within two field ranges associated with the 2d-SC phase. In the lower field range preceding the onset of the intermediate $3/4$-plateau, the 2d-SC phase is characterized by a pronounced field-driven evolution of the local magnetization of the monomer spins, which gradually turns towards the field direction while the local magnetizations of the dimer spins remain nearly unchanged. All local magnetizations are effectively frozen once the system enters the 2d-QFM phase manifested as the intermediate 3/4-plateau. Upon leaving the 2d-QFM phase, the system re-enters the 2d-SC phase in the higher field range, where the total magnetization increases primarily due to a further progressive polarization of the monomer spins complemented by a significant contribution arising from a canting-driven reorientation of the dimer spins located on the vertical diamond units. 

Unlike the previous distorted case, we now turn to the second type of distortion $\delta_{1}=1$, which corresponds to the spin-1/2 Heisenberg antiferromagnet on a distorted diamond-decorated honeycomb lattice where the coupling constant between the monomer and dimer spins on the zigzag diamonds is twice as strong as that on the vertical diamonds, i.e. $J_{1}' = 2J_{1}$. The resulting ground-state phase diagram shown in Fig.~\ref{fig:GSPD}(c) shares several phases already identified in the ground-state phase diagram for the negative value of the distortion parameter $\delta_{1}=-0.5$ including the 2d-QFI, 0d-MD, and 2d-SC phases although their stability regions are quantitatively modified. The most striking difference compared with the regime $\delta_{1}<0$ lies in the complete absence of two gapped quantum ground states manifested in zero-temperature magnetization curves as the intermediate 3/4-plateau. More specifically, the second type of the distortion does not support neither the presence of the 1d-CFM phase in a highly frustrated regime nor the 2d-QFM phase embedded in a less frustrated regime inside of the 2d-SC phase. The finite-size results presented in Fig.~\ref{fig:GSPD}(d) strongly suggest that the intermediate 3/4-plateau completely vanishes from the zero-temperature magnetization curve in the thermodynamic limit for the second type of distortion $\delta_{1}=1$. As a result, the 2d-SC phase developing from the 2d-QFI phase by a continuous field-induced quantum phase transition extends for $\delta_{1}=1$ up to the saturation field. 

Another significant qualitative difference between the ground-state phase diagrams obtained for two representative distortion parameters $\delta_{1}<0$ and $\delta_{1}>0$ can be found in the parameter region with moderate spin frustration at low enough magnetic fields. When the coupling constant between the monomer and dimer spins within the zigzag diamonds $J_{1}'>J_{1}$ (i.e. $\delta_{1}>0$) is stronger than that on the vertical diamonds, the system promotes a regular alternation of singlet tetramers and singlet dimers along the zigzag diamond spin chains, while all vertical diamond units host dimer singlets. This results in a fully fragmented 0d dimer–tetramer liquid (0d-DTL) phase, which differs from the analogous 0d-DTS phase by a massive degeneracy originating from two symmetry-related configurations of each zigzag diamond spin chain having all singlet tetramers tilted either to the left or to the right (cf. the 0d-DTS and 0d-DTL phases in Fig.~\ref{fig:fazy}). The overall degeneracy of the 0d-DTL phase grows exponentially with the linear system size $L$ as $2^{L}$ although the residual entropy per spin remains strictly zero in the thermodynamic limit.

Last but not least, positive values of the distortion parameter $\delta_{1}>0$ stabilize the 1d quantum ferrimagnetic (1d-QFI) phase whose microscopic nature can be inferred from the local magnetizations shown in Fig.~\ref{fig:fazy} for the representative case $\delta_{1}=1$. In the 1d-QFI phase, all vertical diamond units host dimer singlets effectively reducing dimensionality to ferrimagnetic zigzag diamond spin chains. In agreement with this picture, the local magnetization of the dimer spins from the vertical diamonds completely vanishes $m_{3-4}=0$, whereas positive and negative values of the local magnetization $m_{5-8}=0.41$ of the dimer spins from the zigzag diamonds and $m_{1-2}=-0.31$ of the monomer spins reflect the ferrimagnetic character of the zigzag diamond spin chains. It is noteworthy that the 1d-QFI phase is a gapped quantum ground state, which manifests itself in a zero-temperature magnetization curve as the intermediate 1/4-plateau.

Fig.~\ref{fig:GSPD}(d) illustrates a typical zero-temperature magnetization curve for the representative value of the distortion parameter $\delta_{1}=1$ and the particular value of the coupling ratio $J_{2}/J_{1} = 2.5$. Under these conditions, the system first undergoes two discontinuous field-driven phase transitions from the 0d-DTL phase to the 1d-QFI phase and subsequently from the 1d-QFI phase to the 2d-QFI phase. These transitions are manifested as abrupt magnetization jumps between the zero-magnetization plateau (0d-DTL), the $1/4$-plateau (1d-QFI), and the $1/2$-plateau (2d-QFI), respectively. The gapped 2d-QFI phase persists until the external magnetic field closes the spin gap at a continuous field-induced quantum phase transition into the 2d-SC phase. The continuous rise of the magnetization within the field range inherent to the 2d-SC phase is confirmed by a finite-size analysis as indicated by the comparison between ED and DMRG data. This analysis eventually reveals within the respective field range a progressive suppression of the magnetization steps and jumps with increasing system size, as well as, the absence of the gapped 2d-QFM phase embedded as the intermediate 3/4-plateau within the 2d-SC phase predicted previously for 
$\delta_{1}<0$.

\section{Finite-temperature properties}
\label{response}

\begin{figure*}[t!]
	\centering
	\includegraphics[width=1\columnwidth]{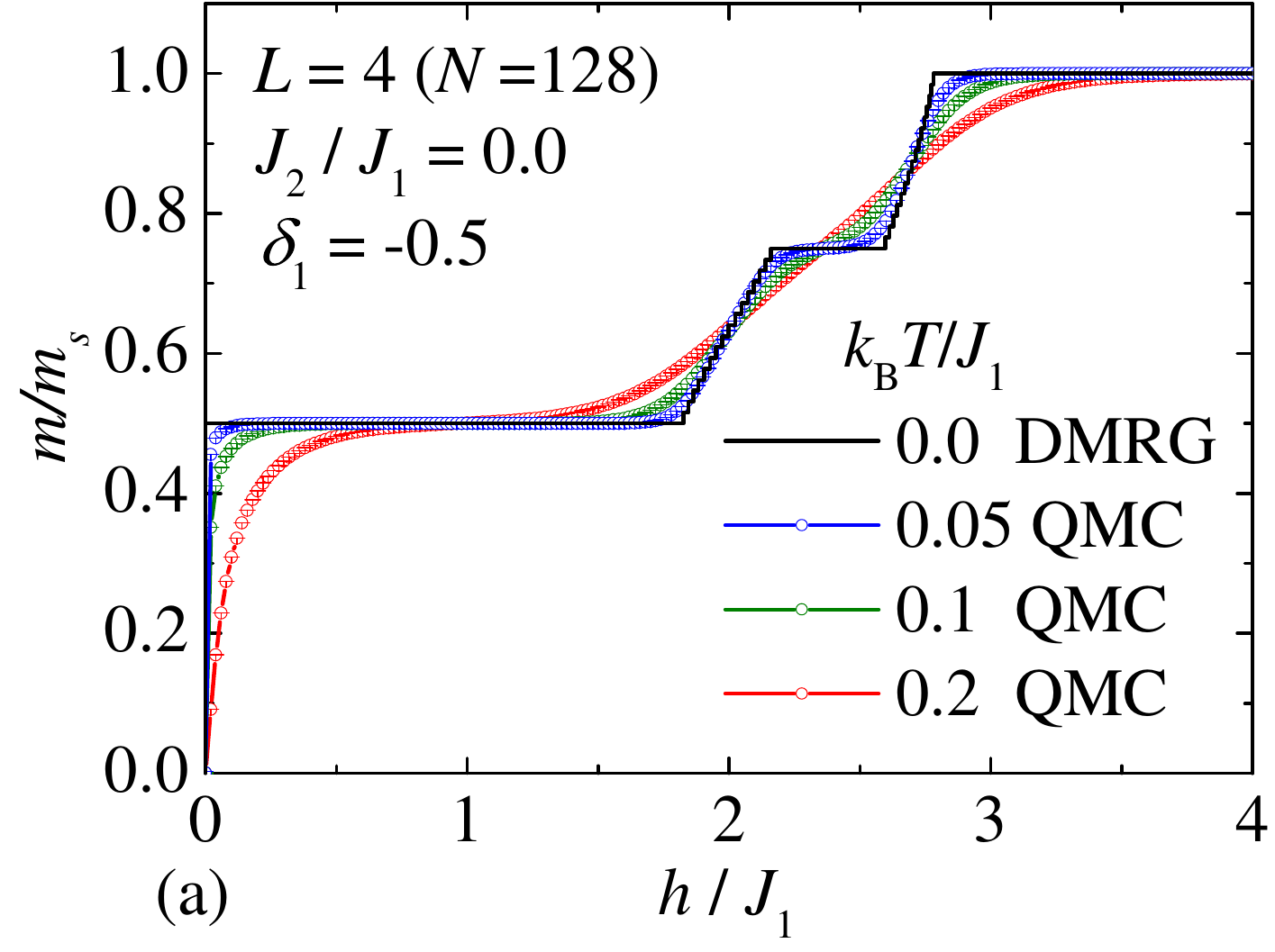}
	\includegraphics[width=1\columnwidth]{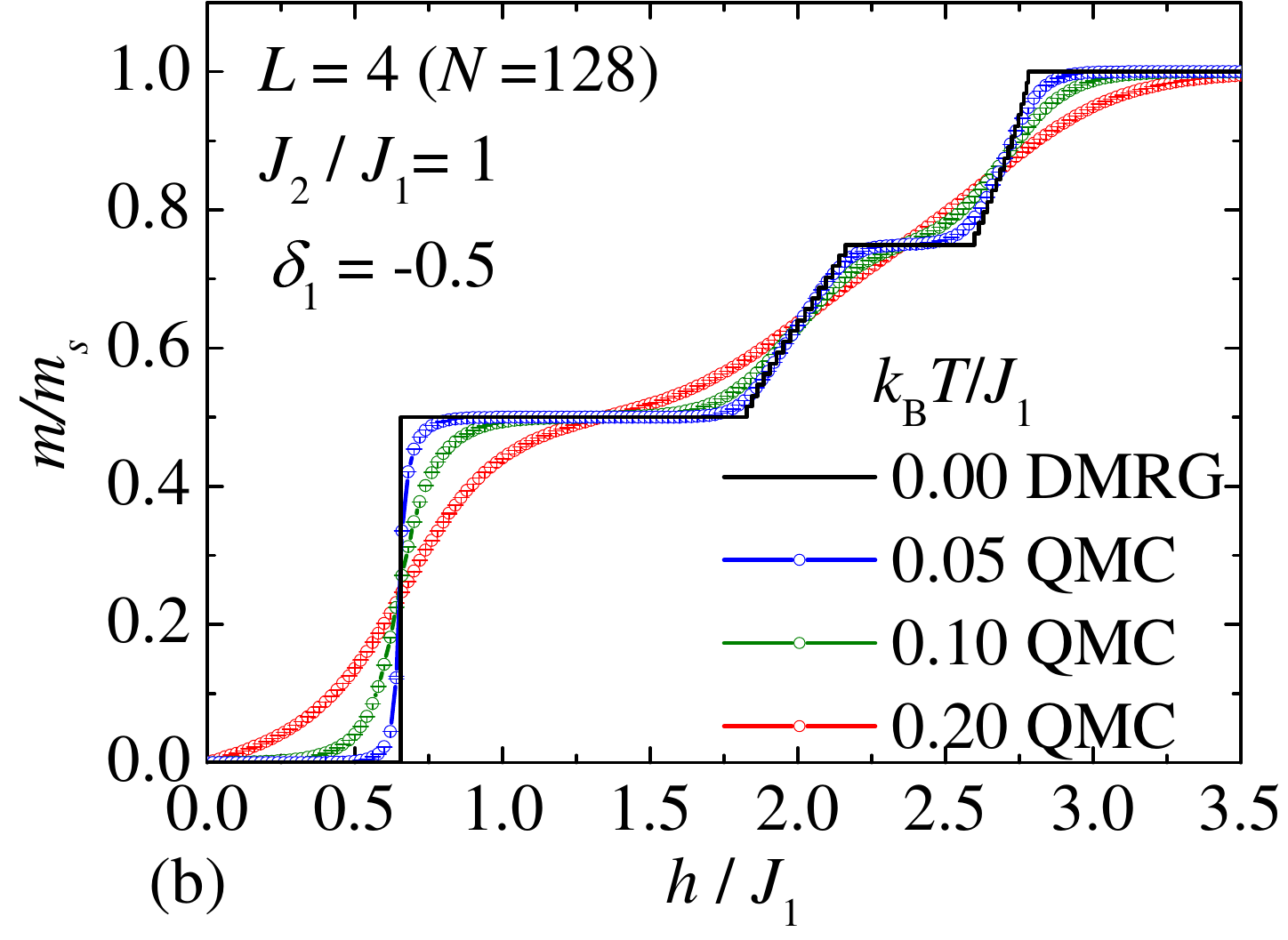}
	\includegraphics[width=1\columnwidth]{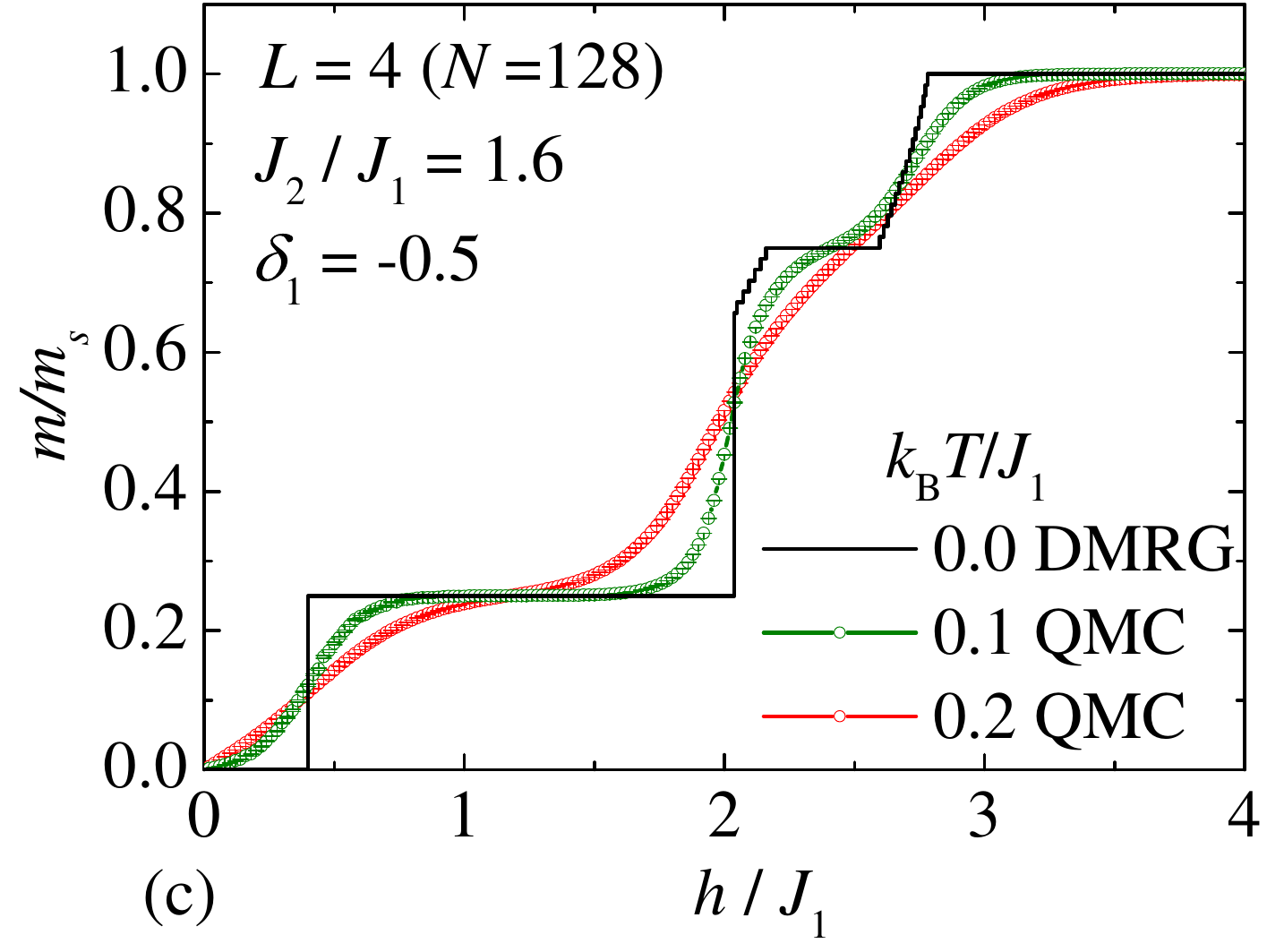}
	\includegraphics[width=1\columnwidth]{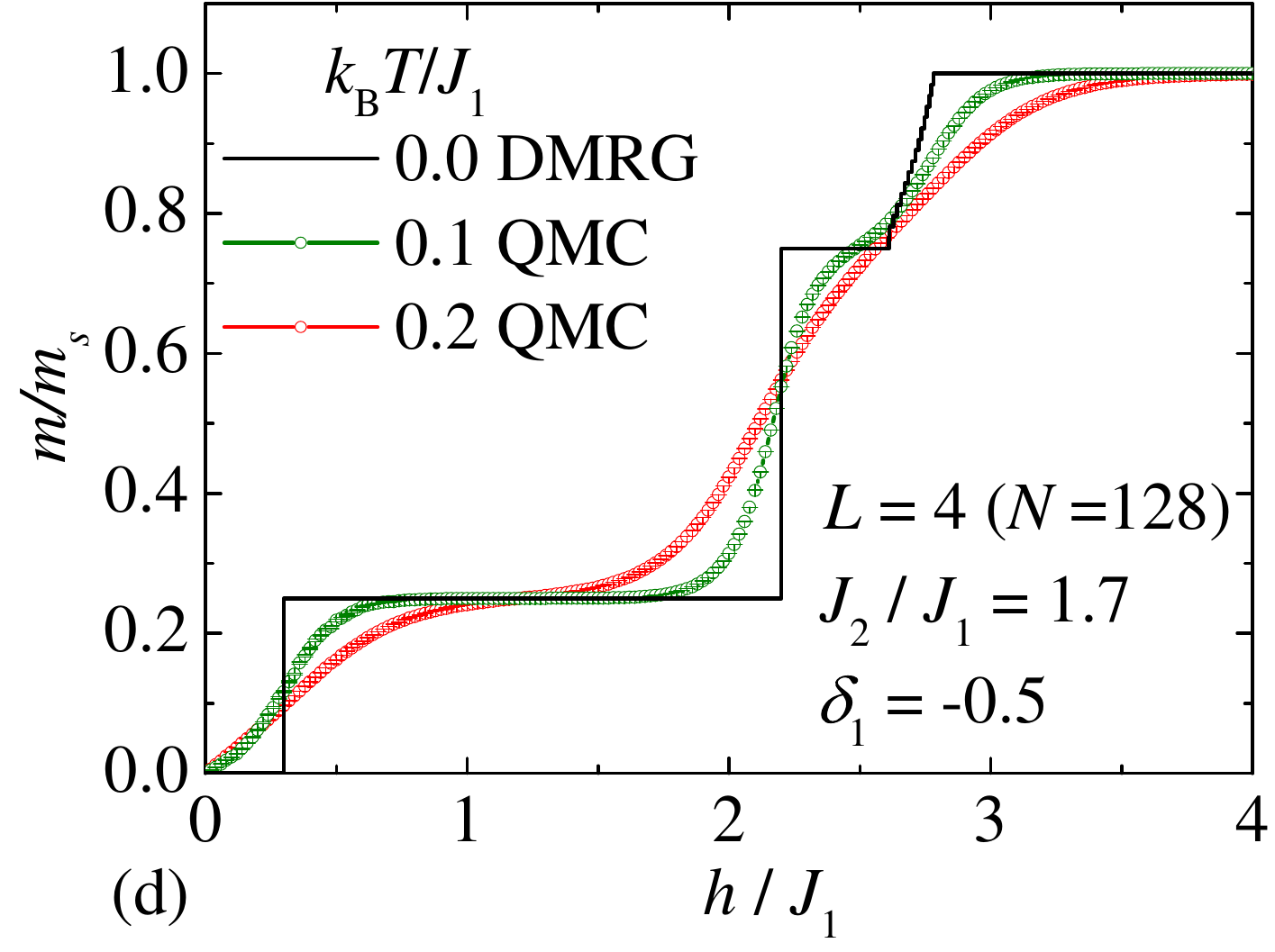}
	\caption{Magnetization curves of the spin-1/2 Heisenberg model on the diamond-decorated honeycomb lattice for distortion parameter $\delta_1=-0.5$. All data correspond to a system of linear size $L=4$ ($N=128$). The zero-temperature DMRG results are compared with finite-temperature QMC simulations at several values of $k_{\rm B}T/J_1$. Panels (a) to (d) show four representative values of the coupling ratio: (a) $J_2/J_1=0$, (b) $J_2/J_1=1$, (c) $J_2/J_1=1.6$, (d) $J_2/J_1=1.7$.}
	\label{fig:mc5}
\end{figure*}

\subsection{Quantum Monte Carlo simulations}

In this section, we examine first finite-temperature magnetization process of the spin-$1/2$ Heisenberg model on a distorted diamond-decorated honeycomb lattice by making use of QMC simulations in a special basis completely avoiding sign problem. In Fig.~\ref{fig:mc5} we display isothermal magnetization curves for one representative value of the negative distortion $\delta_{1}=-0.5$ as obtained from QMC simulations for a system of linear size $L=4$ (i.e. the number of spins $N=128$) at a few selected temperatures, whereas the zero-temperature magnetization curve obtained from DMRG simulations for the same system size serves as a reference. Fig.~\ref{fig:mc5}(a) shows the magnetization curve for the particular value of the interaction ratio $J_{2}/J_{1}=0$, where the intermediate 1/2- and 3/4-plateaus can be associated with the 2d-QFI phase and the 2d-QFM phase, respectively. At lowest temperature $k_{\mathrm{B}}T/J_{1}=0.05$, the magnetization curve from QMC simulations almost perfectly coincides with the zero-temperature magnetization curve obtained from DMRG simulations when exhibiting only a weak thermal broadening at edges of the magnetization plateaus. The robust intermediate 1/2-plateau can be clearly recognized even at much higher temperatures such as $k_{\mathrm{B}}T/J_{1}=0.1$ or $0.2$, while the narrower $3/4$-plateau becomes rapidly smoothed when it turns to an inflection point at the temperature $k_{\mathrm{B}}T/J_{1}=0.1$ or is completely smeared out at the temperature $k_{\mathrm{B}}T/J_{1}=0.2$.

The isothermal magnetization curves shown in Fig.~\ref{fig:mc5}(b) for $J_{2}/J_{1}=1$ exhibit an additional zero-magnetization plateau, which emerges due to the appearance of the 2d-DTS phase. Again, the lowest-temperature QMC data calculated for $k_{\mathrm{B}}T/J_{1}=0.05$ reproduce the DMRG curve with remarkable accuracy including the two field regions associated with the gapless 2d-SC phase, over which the magnetization increases continuously even at absolute zero temperature in the thermodynamic limit and finite temperatures further help to smooth the magnetization curve into a smooth field dependence. Increasing temperature progressively suppresses the zero-magnetization plateau and smears a sharp magnetization jump associated with the discontinuous field-induced transition from the 2d-DTS phase to the 2d-QFI phase.

The magnetization curves calculated for the two intermediate values of the interaction ratio $J_2/J_1=1.6$ and $1.7$ shown in Figs.~\ref{fig:mc5}(c) and \ref{fig:mc5}(d) display several notable features. In both cases, the low-field region exhibits a zero-magnetization plateau originating from the 0d-DTS phase, which is followed by abrupt magnetization jump towards the robust 1/4-plateau associated with the 0d-MD phase. The intermediate 1/4-plateau is quite stable with respect to thermal effects when it remains clearly visible even at elevated temperatures $k_{\rm B}T/J_1=0.1$ and 0.2. At higher magnetic fields, the magnetization curves for these two similar coupling ratios begin to differ. While the system enters for $J_2/J_1=1.6$ the intermediate 3/4-plateau originating from the 2d-QFM phase, the intermediate 3/4-plateau for $J_2/J_1=1.7$ has different microscopic origin as it comes from the 1d-CFM phase. Consequently, the field-induced 2d-SC phase appears for $J_2/J_1=1.7$ only above 
the 3/4-plateau in contrast to the case $J_2/J_1=1.6$ where the critical 2d-SC regime spans both field intervals below and above the 3/4 plateau.

\begin{figure*}[t!]
	\centering
	\includegraphics[width=1\columnwidth]{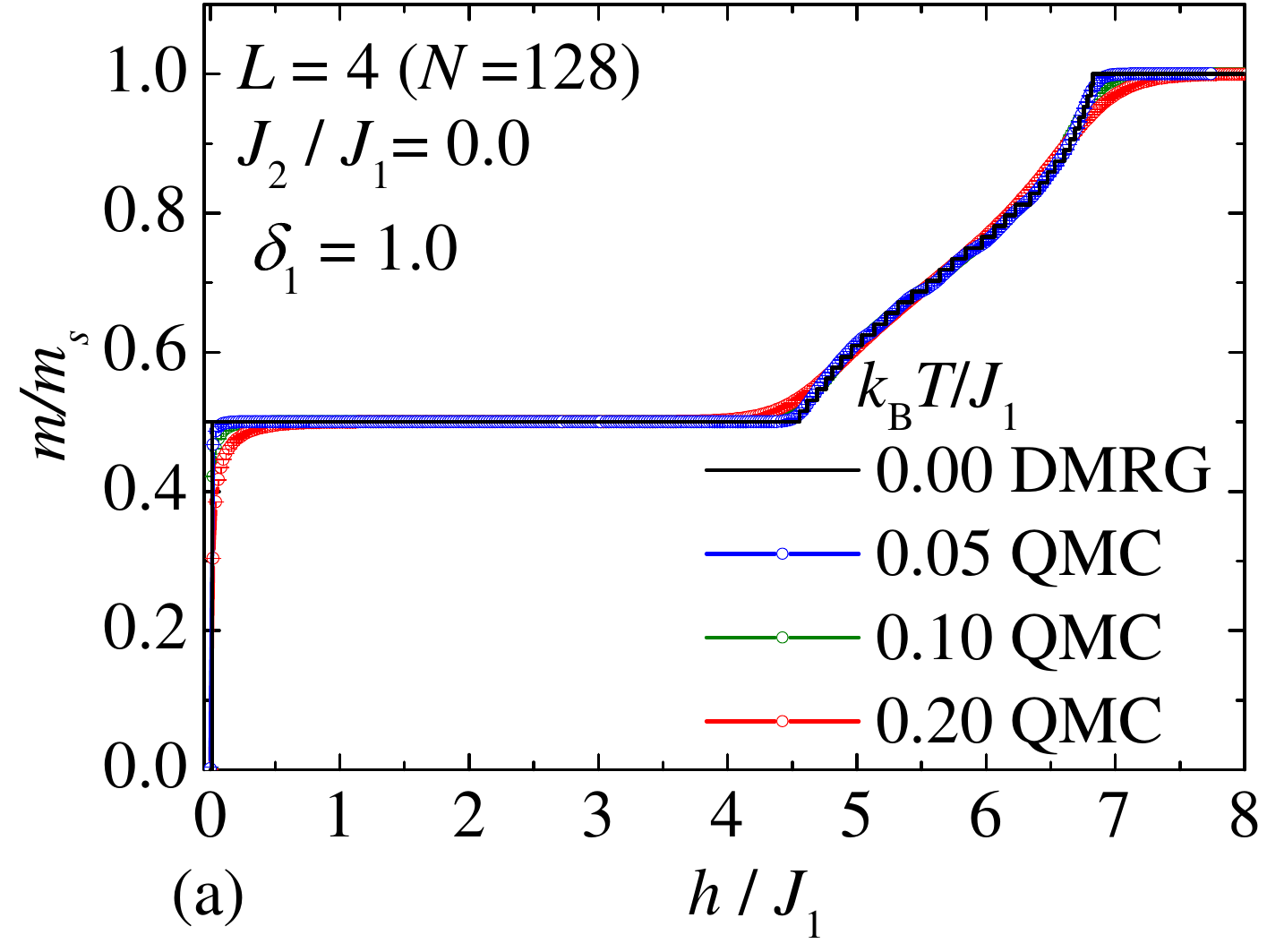}
	\includegraphics[width=1\columnwidth]{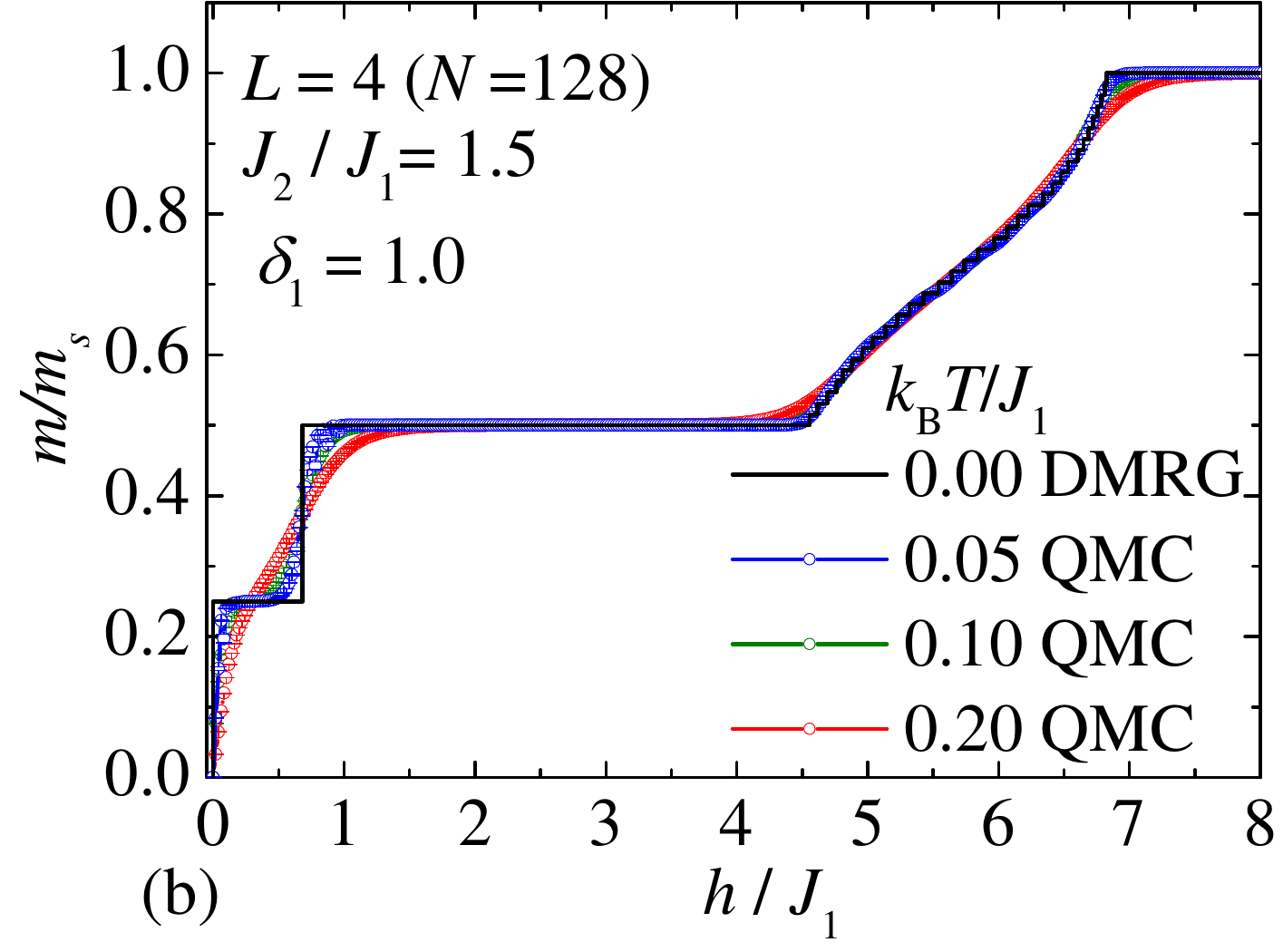}
		\includegraphics[width=1\columnwidth]{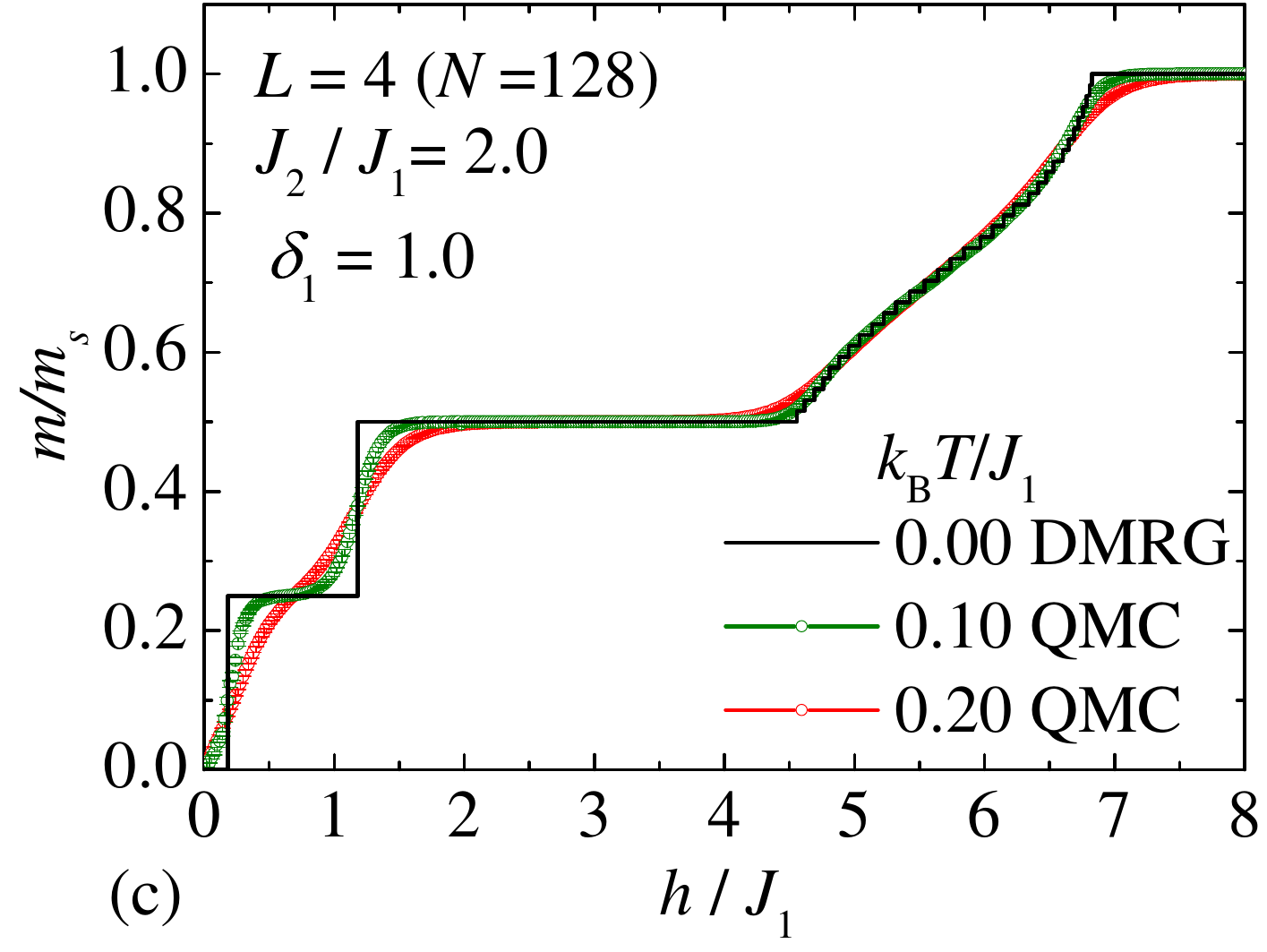}
	\includegraphics[width=1\columnwidth]{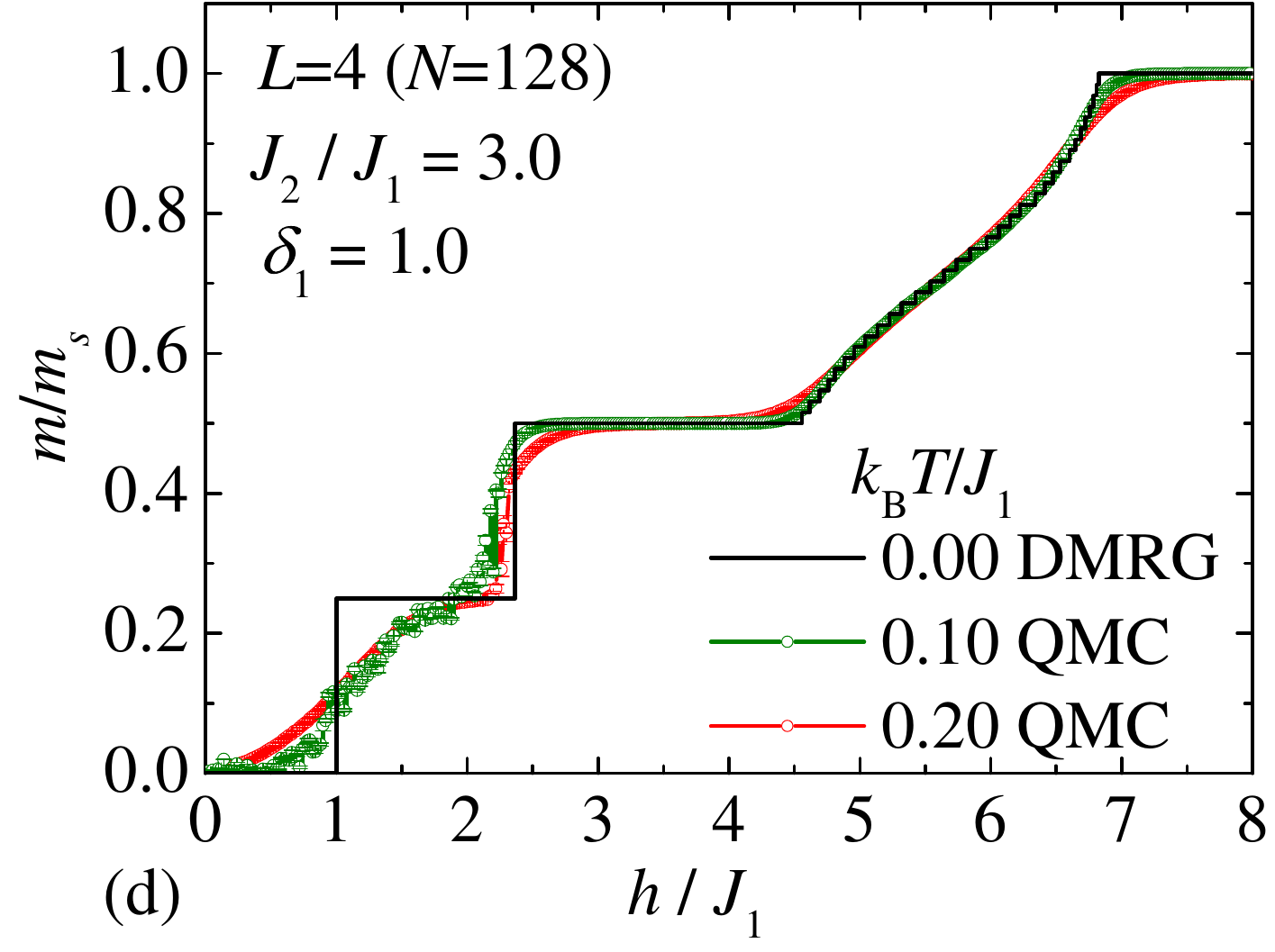}
	\caption{Magnetization curves of the spin-1/2 Heisenberg model on the diamond-decorated honeycomb lattice for distortion parameter $\delta_1=1$. All data correspond to a system of linear size $L=4$ ($N=128$). The zero-temperature DMRG results are compared with finite-temperature QMC simulations at several values of $k_{\rm B}T/J_1$. Panels (a) to (d) show four representative values of the coupling ratio: (a) $J_2/J_1=0$, (b) $J_2/J_1=1.5$, (c) $J_2/J_1=2$, (d) $J_2/J_1=3$.}
	\label{fig:mc1}
\end{figure*}

Next, we examine finite-temperature magnetization curves of the spin-$1/2$ Heisenberg model on a distorted diamond-decorated honeycomb lattice for the second type of distortion $\delta_{1}=1$. To this end, a few isothermal magnetization curves as obtained from QMC simulations for the system size $L=4$ (i.e. $N=128$ spins) are compared in Fig.~\ref{fig:mc1} with the respective zero-temperature magnetization data as obtained from DMRG simulations. The magnetization curves plotted in Fig.~\ref{fig:mc1}(a) for the coupling ratio $J_{2}/J_{1}=0$ display the characteristic 1/2-plateau originating from the 2d-QFI phase, which end up at a continuous field-driven phase transition into the 2d-SC phase with continuously varying magnetization eventually terminating at the saturation field. The QMC data for the low-temperature magnetization curves are in an excellent agreement with the zero-temperature magnetization curve obtained from DMRG simulations with only a small thermal rounding effect observable close to both field-induced quantum phase transitions. Among other matters, the QMC results provide an independent confirmation for the absence of the intermediate 3/4-plateau associated with the 2d-QFM phase.

The magnetization curves depicted in Fig.~\ref{fig:mc1}(b) for $J_{2}/J_{1}=1.5$ initially enters the narrow 1/4-plateau associated with the 1d-QFI phase before the system again successively enters 
the 2d-QFI and 2d-SC phases at higher magnetic fields. Despite its relatively small width, the intermediate 1/4-plateau can be clearly discerned in the magnetization curves computed at two lowest temperatures $k_{\rm B}T/J_{1}=0.05$ and $0.10$. It is evident from Fig.~\ref{fig:mc1}(b) that thermal fluctuations progressively smear first the narrow 1/4-plateau, then the robust 1/2-plateau, whereas the continuous rise in the magnetization within the 2d-SC phase has the highest resistance against the thermal smoothing.

The low-temperature magnetization curve displayed in Fig.~\ref{fig:mc1}(c) for the interaction ratio $J_{2}/J_{1}=2.0$ is indicative of a tiny zero-magnetization plateau, which originates from the 0d-DTL ground state at very low magnetic fields. Although this plateau is extremely narrow, it remains visible at the lowest simulated temperature $k_{\rm B}T/J_{1}=0.1$. The overall magnetization profile including a sequence of intermediate plateaus emergent at higher fields mirrors that in Fig.~\ref{fig:mc1}(b). A width of the zero-magnetization plateau associated with the 0d-DTL phase can be 
noticeably enhanced by increasing the interaction ratio as demonstrated by the magnetization curve plotted for $J_{2}/J_{1}=3$ in Fig.~\ref{fig:mc1}(d). The QMC data for the magnetization curve calculated at the lowest temperature $k_{\rm B}T/J_{1}=0.10$ again confirm presence of 0-, 1/4-, and 1/2-plateaus associated with the 0d-DTL, 0d-MD, and 2d-QFI phases before the systems enters a broad field region corresponding to the 2d-SC phase terminating only at the saturation field.

\subsection{Effective lattice-gas description for $\delta_1<0$}

Flat one-magnon bands often play a central role in determining the low-energy spectrum of geometrically frustrated quantum magnets. A completely dispersionless band implies the existence of a strictly localized magnon, which in turn enables the construction of many-magnon eigenstates from independent bound one-magnon states. A set of the localized many-magnon eigenstates can be naturally described within effective lattice-gas models, the concept pioneered and extensively developed  by Johannes Richter and collaborators \cite{rich02,derz06,derz07,derz11,derz15}. In this part, we develop an effective lattice-gas model capable of capturing the low-temperature features of the spin-$1/2$ Heisenberg model on a diamond-decorated honeycomb lattice for negative values of the distortion parameter as demonstrated by the representative case with $\delta_1=-0.5$. It follows from the ground-state phase diagram shown in Fig. \ref{fig:GSPD}(a) that the system host in the highly frustrated regime $J_2/J_1 \gtrsim 1.8$ three fragmented 0d-DTS, 0d-MD, and 1d-CFM phases, all of which can be generated  from the fully polarized ferromagnetic state by introducing local dimer and/or tetramer singlets. 

The effective lattice-gas model can be formulated by assigning three types hard-core objects to the relevant local eigenstates: dimer singlets on vertical diamond units, dimer singlets on zigzag diamond units, and tetramer singlets on vertical diamond units. The dimer singlets are represented by hard-core particles occupying decorating sites of a Lieb-type honeycomb lattice, while the tetramer singlets are represented by hard-core particles extending over the bonds of this lattice. The chemical potentials $\mu_1 = J_{1}+J_{2}-h$, $\mu_2 = J_{1}'+J_{2}-h$ and $\mu_3 = 3J_{1}-2h$ assigned to the three distinct hard-core particles account for the energy cost associated with creating a dimer singlet on a vertical diamond unit, a dimer singlet on a zigzag diamond unit, and a tetramer singlet on a vertical diamond unit, respectively. The monomer spins $S_{1,i}$ and $S_{2,i}$ remain free unless they are involved in a tetramer singlet on a vertical diamond unit. Within this description, the effective lattice–gas model can be defined by the Hamiltonian
\begin{eqnarray}
{\cal H}_{\rm eff}&=&E_{\rm FM}^{(0)}-h\sum_{i=1}^{\cal N} \left[3+S_{1,i}^z(1-d_{v,i})+S_{2,i}^z(1-d_{v,i})\right] \nonumber \\
&-&\mu_1\sum_{i=1}^{\cal N} (n_{l,i}+n_{r,i})-\mu_2\sum_{i=1}^{\cal N} n_{v,i}-\mu_3\sum_{i=1}^{\cal N} d_{v,i},
\label{lgham}
\end{eqnarray}
where $E_{\rm FM}^{(0)} = \frac{3}{4} {\cal N} J_2 +2 {\cal N} J_1'+{\cal N} J_1$ denotes the zero-field energy of the fully polarized ferromagnetic state serving as a reference state and ${\cal N}$ is the total number of unit cells. The occupation numbers $n_{l,i}, n_{r,i}, n_{v,i}\in{\{0,1\}}$ determine absence or presence of a dimer singlet on the left–, right- or vertically-oriented diamond unit cell, while the occupation number $d_{v,i}\in{\{0,1\}}$ determines absence or presence of a tetramer singlet on a vertical diamond unit cell. 

The partition function of the effective lattice–gas model defined by the Hamiltonian (\ref{lgham}) is given by
\begin{eqnarray}
{\cal Z}\!&=&\! \exp(-\beta E_{\rm FM}^{(0)}\!+\!3\beta {\cal N} h)\sum_{\{S_{1,i}^z\}}\!\sum_{\{S_{2,i}^z\}}\!\sum_{\{n_{l,i}\}}\!\sum_{\{n_{r,i}\}}\!\sum_{\{n_{v,i}\}}\!\sum_{\{d_{v,i}\}}\nonumber \\
&\times&\!\!\prod_{i}\frac{1}{4^{d_{v,i}}}(1\!-\!n_{v,i}d_{v,i})\exp\!\left[\!\beta h \sum_{i=1}^{\cal N}\!(S_{1,i}^z\!+\!S_{2,i}^z)(1\!-\!d_{v,i})\!\right] \nonumber \\
&\times&\!\exp\!\left[\beta\mu_1\!\sum_{i=1}^{\cal N}(n_{l,i}\!+\!n_{r,i})\!+\!\beta\mu_2\!\sum_{i=1}^{\cal N} \!n_{v,i}\!+\!\beta\mu_3\!\!\sum_{i=1}^{\cal N} \!d_{v,i}\right]\!.
\end{eqnarray}
where the summations run over all possible configurations of the monomer spins ($S_{1,i}^{z}$ and $S_{2,i}^{z}$) and the occupation numbers ($n_{l,i}$, $n_{r,i}$, $n_{v,i}$, and $d_{v,i}$) encoding presence or absence of the hard-core particles. The factor $(1 - n_{v,i} d_{v,i})$ serves as a local projection operator enforcing the hard–core constraint on each vertical diamond unit, which cannot simultaneously host both a dimer and a tetramer singlet. The prefactor $1/4^{d_{v,i}}$ compensates for the artificial fourfold degeneracy that would otherwise arise when summing over states of monomer spins provided that they belong to a singlet tetramer. 

After performing the summations over all available configurations of the monomer spins and the hard–core objects, the partition function of the effective lattice–gas model takes the closed analytic form
\begin{eqnarray}
{\cal Z}&=& \exp(-\beta E_{\rm FM}^{(0)}\!+\!3\beta {\cal N} h)\!\left[1\!+\!\exp(\beta\mu_2)\right]^{2{\cal N}}\!\!\left[\!2\cosh\!\!\left(\!\!\frac{\beta h}{2}\!\!\right)\!\right]^{\!2{\cal N}}\nonumber \\ &\times&\left[1+\exp(\beta\mu_1)+\exp(\beta\mu_3)\right]^{{\cal N}}.
\end{eqnarray}
The corresponding free energy per unit cell $f =-\frac{1}{{\cal N}} k_{\rm B}T\ln{\cal Z}$ is obtained in the form
\begin{eqnarray}
f \!&=&\! \frac{3}{4}J_2 +2J_1'+J_1-3h-\frac{2}{\beta}\ln\left[1+\exp(\beta\mu_2)\right] \nonumber \\
\!&-&\! \frac{1}{\beta}\ln\!\left\{\!\!\left[2\cosh\!\!\left(\frac{\beta h}{2}\right)\!\right]^2\!\!\left[1\!+\!\exp(\beta\mu_1)\!+\!\exp(\beta\mu_3)\right]\right\}\!.
\end{eqnarray}
The resulting free energy provides a direct access to the thermodynamic quantities of interest, in particular, the magnetization is obtained straightforwardly from
\begin{equation}
m = -\frac{\partial f}{\partial h}.
\end{equation}

\begin{figure}[t!]
	\centering
	\includegraphics[width=1\columnwidth]{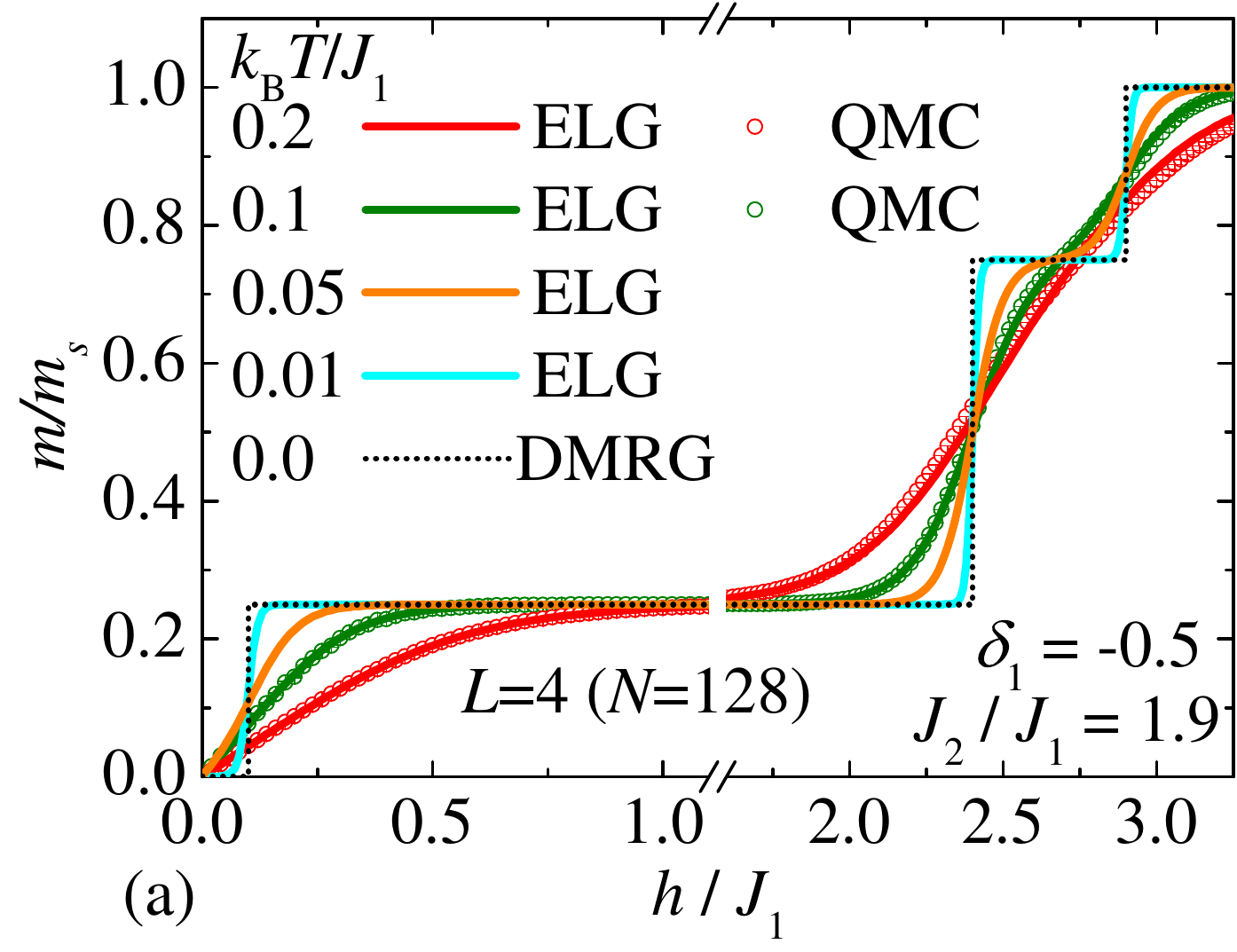}
	\includegraphics[width=1\columnwidth]{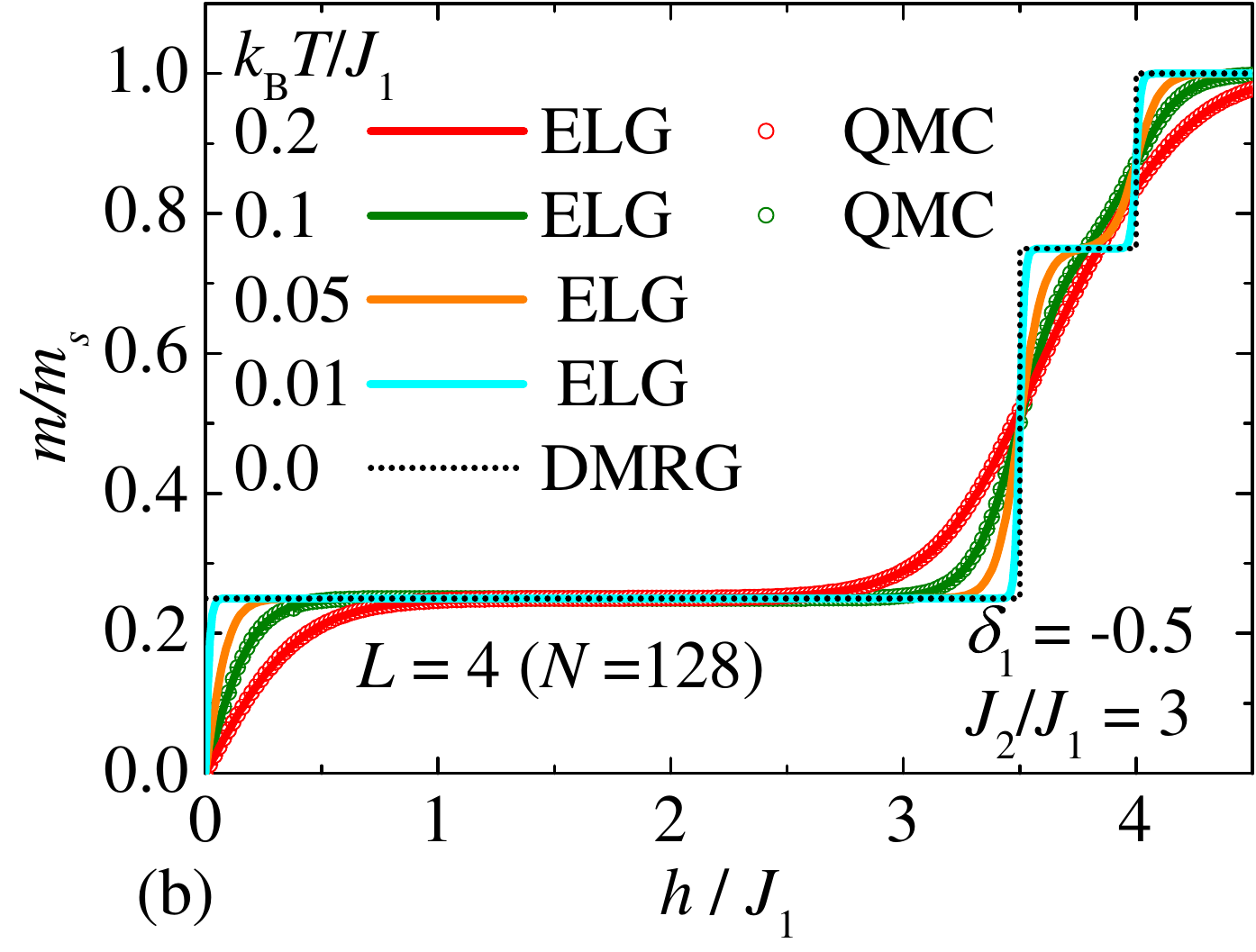}
	\caption{Magnetization curves of the spin-$\frac{1}{2}$ Heisenberg model on the diamond-decorated honeycomb lattice for the fixed distortion parameter $\delta_{1}=-0.5$ and two distinct values of the coupling ratio: (a) $J_{2}/J_{1}=1.9$; (b) $J_{2}/J_{1}=3$. The zero-temperature DMRG data for a system of size $L=4$ ($N=128$) are shown by black dotted lines. Finite-temperature results obtained from QMC simulations (symbols) for the same system size are compared with analytical predictions of the ELG model (solid curves).}
	\label{fig:ELG}
\end{figure}

In Fig.~\ref{fig:ELG} we directly compare finite-temperature magnetization curves obtained from QMC simulations with the analytical predictions calculated according to the effective lattice-gas (ELG) description for the specific value of the distortion parameter $\delta_{1}=-0.5$. A zero-temperature magnetization curve as obtained from DMRG simulations is also plotted for comparison. Panels (a) and (b) compare isothermal magnetization curves as obtained from the QMC and ELG approaches for two representative values of the coupling ratio $J_{2}/J_{1}=1.9$ and $3.0$. Both investigated cases show a characteristic sequence of the fractional magnetization plateaus: the intermediate 0-, 1/4-, and 1/2-plateaus observed for the lower value of the interaction ratio $J_{2}/J_{1}=1.9$ correspond to the 0d-DTS, 0d-MD, and 1d-CFM phases, while only the latter two intermediate 1/4-, and 1/2-plateaus stemming from the 0d-MD and 1d-CFM phases persist for the higher value of the interaction ratio 
$J_{2}/J_{1}=3$ for which zero-plateau vanishes.

It can be found from Fig.~\ref{fig:ELG} that the QMC data follow the ELG predictions with nearly perfect agreement over the entire field range at the lower selected temperature $k_{\rm B}T/J_{1}=0.1$, at which thermal fluctuations noticeably smooth the step-like profile of the zero-temperature magnetization curve. Even at moderate temperature $k_{\rm B}T/J_{1}=0.2$, the ELG predictions remain quantitatively accurate and capture all essential features of QMC data. A key advantage of the ELG description is its ability to access very low temperatures beyond the reach of unbiased QMC sampling.  
Because QMC simulations of frustrated quantum magnets encounter severe ergodicity and freezing issues at low temperatures, particularly in the highly frustrated region, reliable data below $k_{\rm B}T/J_{1}\approx 0.05$ are generally not accessible. In contrast, the ELG modeling remains straightforward to evaluate and yields smooth well-behaved magnetization curves at arbitrarily low temperatures.  
These results demonstrate  that the effective lattice-gas framework faithfully reproduces the essential low-temperature physics of the model and effectively complements the QMC simulations in the parameter regime where numerical sampling becomes inefficient.

\section{Conclusion}
\label{conclusion}
In this work, we have presented a comprehensive study of the spin-1/2 Heisenberg antiferromagnet on the distorted diamond-decorated honeycomb lattice focusing on how lattice distortion and magnetic field jointly shape its rich quantum behavior. By combining large-scale DMRG, Lanczos-algorithm-based ED, sign-problem-free QMC simulations, and an analytically tractable effective lattice-gas approach, we have mapped out the ground-state phase diagrams for representative distortions and uncovered a remarkable diversity of quantum phases. These include 2d quantum ferrimagnetic, quantum ferromagnetic, and spin-canted states, fragmented 0d monomer–dimer state, 0d dimer–tetramer solid, and 0d dimer-tetramer liquid, as well as, 1d classical ferromagnetic and quantum ferrimagnetic phases, which give rise to several field-induced phase transitions and manifest themselves as fractional magnetization plateaus.

The interplay between geometric frustration, local conservation laws, and anisotropic coupling hierarchies results in a sequence of discontinuous and continuous field-driven transitions, directly reflected in the structure of the magnetization curves. Negative distortion stabilizes a set of fragmented phases that can be naturally interpreted in terms of localized-magnon physics, enabling us to construct an effective lattice-gas description that reproduces the QMC results with high accuracy even in the challenging low-temperature regime. Positive distortion, by contrast, suppresses the gapful states associated with the intermediate 3/4-plateau and promotes 1d quantum ferrimagnetic order and a massively degenerate dimer–tetramer liquid.

Beyond the results explored here, our findings indicate that the distorted diamond-decorated lattice hosts even richer physics than captured within this magnetic-field study. In particular, the structure of the degenerate dimer–tetramer manifold closely mirrors the constrained sectors known to generate Kasteleyn-type string physics in related geometries. In a separate work on the $J_2$-distorted version of the same model \cite{ensamble}, we demonstrate that such constraints can drive a pronounced approach to the Kasteleyn-type phase transition. The close correspondence between the constrained sectors of both distortion patterns suggests that similarly unconventional mechanisms—such as string proliferation, asymmetric thermodynamic responses, and emergent classical criticality—may also arise in the $J_1$-distorted model studied here, albeit in a parameter regime beyond the scope of the present work.

Altogether, these observations establish the distorted diamond-decorated honeycomb lattice as a versatile platform where localized-magnon physics, dimensional reduction, and emergent constrained phenomena intertwine. They further open a natural route toward uncovering Kasteleyn-related mechanisms and other unconventional collective behaviors in two-dimensional frustrated quantum magnets, some of which will be elaborated in future works.

\begin{acknowledgments}
The authors dedicate the present paper to the memory of Johannes Richter, whose inspiring discussions, insightful suggestions, and valuable comments significantly shaped their research interests.  
KK would like to thank Stefan Wessel and Nils Caci for providing the QMC code developed by Lukas Weber, for their guidance concerning implementation of QMC simulations for frustrated quantum spin systems, and for helpful consultations during the initial phase of this project. Computing time for the QMC simulations was provided by the IT Center at RWTH Aachen University.
\end{acknowledgments}

\section*{Research funding:}
Funded by the EU NextGenerationEU through the Recovery and Resilience Plan for Slovakia under the project No. 09I03-03-V04-00403. 

\section*{Author contributions:}
Both authors contributed equally to this work.

\section*{Use of Large Language Models, AI and Machine Learning Tools:}
After finalizing the manuscript ChatGPT tool was used to improve the language correctness.

\section*{Conflict of interest:}
The authors declare that there are no conflicts of interest regarding the publication of this article.

\section*{Research ethics:}
This research involves no human participants, no animal subjects, and no biological materials. Therefore, ethical approval is not required.

\section*{Informed consent:}
Not applicable.

\section*{Data availability:}
All data reported in this study can be obtained from the corresponding author upon reasonable request.


\begin{thebibliography}{100}

\bibitem{diep04} 
H.~T. Diep, 
\textit{Frustrated Spin Systems}, 
World Scientific, Singapore, 2004.

\bibitem{lacr11} 
C. Lacroix, F. Mila, and Ph. Mendels, 
\textit{Introduction to Frustrated Magnetism}, 
Springer, Heidelberg, 2011.

\bibitem{bale10} L. Balents, Spin liquids in frustrated magnets, Nature
\textbf{464}, 199 (2010).

\bibitem{stre17}  J. Stre\v{c}ka, J. Richter, O. Derzhko, T. Verkholyak, and K. Karl'ov\'a,  Diversity of quantum ground states and quantum phase transitions of a spin-1/2 Heisenberg octahedral chain, Phys. Rev. B {\bf 95}, 224415 (2017).

\bibitem{stre22} J. Stre\v{c}ka, T. Verkholyak, J. Richter, K. Karl'ov\'a, O. Derzhko, and J. Schnack, Frustrated magnetism of spin-1/2 Heisenberg diamond and octahedral chains as a statistical-mechanical monomer-dimer problem, Phys. Rev. B \textbf{105}, 064420 (2022).

\bibitem{karl22} 
K. Karl'ov\'a, J. Stre\v{c}ka, and J. Richter, 
Towards lattice-gas description of low-temperature properties above the Haldane and cluster-based Haldane ground states of a mixed spin-(1,1/2) Heisenberg octahedral chain, 
Phys. Rev. E \textbf{106}, 014107 (2022).

\bibitem{hone04} A. Honecker, J. Schulenburg, and J. Richter, 
Magnetization plateaus in frustrated antiferromagnetic quantum spin models, 
J. Phys.: Condens. Matter \textbf{16}, S749 (2004). 

\bibitem{rich04}
J. Richter, J. Schulenburg, A. Honecker, J. Schnack, and H.-J. Schmidt,
Exact eigenstates and macroscopic magnetization jumps in strongly frustrated spin lattices,
J. Phys.: Condens. Matter \textbf{16}, S779 (2004).

\bibitem{rich02} 
J. Schulenburg, A. Honecker, J. Schnack, J. Richter, and H.-J. Schmidt, 
Macroscopic Magnetization Jumps due to Independent Magnons in Frustrated Quantum Spin Lattices, 
Phys. Rev. Lett. {\bf 88}, 167207 (2002).

\bibitem{signProblem}
P. Henelius and A. W. Sandvik, Sign problem in Monte
Carlo simulations of frustrated quantum spin systems,
Phys. Rev. B \textbf{62}, 1102 (2000).

\bibitem{sign2}
D. Hangleiter, I. Roth, D. Nagaj, and J. Eisert, Easing
the Monte Carlo sign problem, Science Advances \textbf{6},
eabb8341 (2020).

\bibitem{schn18}
J. Schnack, J. Schulenburg, and J. Richter, Magnetism of the $N$ = 42 kagome lattice antiferromagnet, Phys. Rev. B \textbf{98}, 094423 (2018).

\bibitem{derz06} 
O. Derzhko and J. Richter, 
Universal low-temperature behavior of frustrated quantum antiferromagnets in the vicinity of the saturation field,
Eur. Phys. J. B \textbf{52}, 23 (2006).

\bibitem{derz07} 
O. Derzhko, J. Richter, A. Honecker, and H.-J. Schmidt,
Universal properties of highly frustrated quantum magnets in strong magnetic fields, 
Low Temp. Phys. \textbf{33}, 745 (2007).

\bibitem{derz11} 
O. V. Derzhko, T. E. Krokhmalskii, and J. Richter,
Quantum Heisenberg antiferromagnet on lowdimensional frustrated lattices, 
Theor. Math. Phys. \textbf{168}, 1236 (2011).

\bibitem{derz15} 
O. Derzhko, J. Richter, and M. Maksymenko,  
Strongly correlated flat-band systems: The route from Heisenberg spins to Hubbard electrons, 
Int. J. Mod. Phys. B \textbf{29}, 1530007 (2015). 

\bibitem{zhit05} 
M.E. Zhitomirsky, H. Tsunetsugu, 
High field properties of geometrically frustrated magnets, 
Prog. Theor. Phys. Suppl. \textbf{160}, 361 (2005).

\bibitem{zhit04}	
M.E. Zhitomirsky, H. Tsunetsugu, 
Exact low-temperature behavior of a kagom\'e antiferromagnet at high fields, 
Phys. Rev. B \textbf{70},  100403 (2004).

\bibitem{okum19}
R. Okuma, D. Nakamura, T. Okubo, A. Miyake, A. Matsuo, K. Kindo, M. Tokunaga, N. Kawashima, S. Takeyama, and Z. Hiroi, 
A series of magnon crystals appearing under ultrahigh magnetic fields in a kagom\'e antiferromagnet,
Nat. Commun. \textbf{10}, 1229 (2019).

\bibitem{webe1} A. Honecker, L. Weber, P. Corboz, F. Mila and S. Wessel,
Quantum Monte Carlo simulations of highly frustrated magnets in a cluster basis: The two-dimensional Shastry-Sutherland model,  J. Phys.: Conf. Ser. \textbf{2207}, 012032 (2022).

\bibitem{alet16} F. Alet, K. Damle, and S. Pujari, Sign-problem-free
Monte Carlo simulation of certain frustrated quantum
magnets, Phys. Rev. Lett. \textbf{ 117}, 197203 (2016).

\bibitem{stap18} J. Stapmanns, P. Corboz, F. Mila, A. Honecker, B. Normand, and S. Wessel,  Thermal Critical Points and Quantum Critical End Point  in the Frustrated Bilayer Heisenberg Antiferromagnet,  Phys. Rev. Lett. \textbf{121}, 127201 (2018).

\bibitem{webe2} L. Weber, A. Honecker, B. Normand, P. Corboz, F. Mila, S. Wessel, Quantum Monte Carlo simulations in the trimer basis: first-order transitions and thermal critical points in frustrated trilayer magnets,
SciPost Phys. \textbf{12}, 054 (2022). 

\bibitem{webe3} L. Weber, N. Caci, and S. Wessel, 
Cluster quantum Monte Carlo study of two-dimensional weakly coupled frustrated trimer antiferromagnets,
Phys. Rev. B \textbf{106}, 035141 (2022).

\bibitem{caci23} N. Caci, K. Karl'ov\'a, T. Verkholyak, J. Stre\v{c}ka, S. Wessel, A. Honecker, Phases of the spin-1/2 Heisenberg antiferromagnet on the diamond-decorated square lattice in a magnetic field
 Phys. Rev. B \textbf{107}, 115143 (2023).

\bibitem{karl24} K. Karl'ov\'a, A. Honecker, N. Caci, S. Wessel, J. Stre\v{c}ka, T. Verkholyak, Thermodynamic properties of the macroscopically degenerate tetramer-dimer phase of the spin-1/2 Heisenberg model on the diamond-decorated square lattice, Phys. Rev. B \textbf{110}, 214429 (2024).

\bibitem{mori16} K. Morita and N. Shibata, Exact nonmagnetic ground
state and residual entropy of S = 1/2 Heisenberg diamond
spin lattices, J. Phys. Soc. Jpn. \textbf{85}, 033705 (2016).

\bibitem{hiro16} Y. Hirose, A. Oguchi, and Y. Fukumoto, Exact realization
of a quantum-dimer model in Heisenberg antiferromagnets
on a diamond-like decorated lattice, J. Phys. Soc. Jpn. \textbf{85}, 094002 (2016).

\bibitem{hiro17} Y. Hirose, S. Miura, C. Yasuda, and Y. Fukumoto, Notes on ground-state properties of mixed spin-1 and spin-1/2 Lieb-lattice Heisenberg antiferromagnets, 
J. Phys. Soc. Jpn. \textbf{86}, 083705 (2017).

\bibitem{hiro18} Y. Hirose, S. Miura, C. Yasuda, and Y. Fukumoto,
Ground-state properties of spin-1/2 Heisenberg antiferromagnets with frustration on the diamond-like-decorated square and triangular lattices, AIP Adv. \textbf{8}, 101427 (2018).

\bibitem{hiro20} Y. Hirose, A. Oguchi, and Y. Fukumoto, Quantum dimer model containing Rokhsar-Kivelson point expressed by spin-1/2 Heisenberg antiferromagnets, Phys. Rev. B \textbf{101}, 174440 (2020).

\bibitem{dmit1}	D.V. Dmitriev,  V.Y. Krivnov,  O.A. Vasilyev,
Macroscopic ground state degeneracy of the Heisenberg model with ferromagnetic and antiferromagnetic interactions on diamond-decorated lattices, Phys. Rev. B \textbf{112}, 094426 (2025).

\bibitem{zhang2000} H.-X. Zhang, Y.-X. Tong, Z.-N. Chen, K.-B. Yu, B.-S. Kang, 
Cyano-bridged extended heteronuclear supramolecular architectures with hexacyanoferrates(II) as building blocks, J. Organomet. Chem. \textbf{598}, 63 (2000).

\bibitem{travnicek2001} Z. Tr\'avn\'a\v{c}ek, Z. Sm\'ekal, A. Escuer, J. Marek, Synthesis, structure and magnetic behaviour of a two-dimensional cyano-bridged complex [\{Cu(ept)\}$_3$Fe(CN)$_6$](ClO$_4$)$_2$ $\cdot$ 5H$_2$O [ept = N-(2-aminoethyl)-1,3-diaminopropane], New J. Chem. \textbf{25}, 655 (2001).

\bibitem{schu02} 
J. Schulenburg and J. Richter, 
Infinite series of magnetization plateaus in the frustrated dimer-plaquette chain,
Phys. Rev. B {\bf 65}, 054420 (2002).

\bibitem{baue11} 
B. Bauer, L.D. Carr, H.G. Evertz, A. Feiguin, J. Freire, S. Fuchs, L. Gamper, J. Gukelberger, E. Gull, S. Guertler, A. Hehn, R. Igarashi, S.V. Isakov, D. Koop, P.N. Ma, P. Mates, H. Matsuo, O. Parcollet, G. Pawlowski, J.D. Picon, L. Pollet, E. Santos, V.W. Scarola, U. Schollw\"ock, C. Silva, B. Surer, S. Todo, S. Trebst, M. Troyer, M.L. Wall, P. Werner, and S. Wessel, 
The ALPS project release 2.0: open source software for strongly correlated systems,
J. Stat. Mech.: Theor. Exp. \textbf{2011}, P05001 (2011).

\bibitem{ensamble}
K. Karl’ová, A. Rufino, T. Verkholyak, N. Caci, S. Wessel, J. Strečka, F. Mila, A. Honecker,
Approaching Kasteleyn transition in frustrated quantum Heisenberg antiferromagnets, preprint arXiv: 2601.14382.

\end{thebibliography}
\end{document}